\documentclass[times,final]{elsarticle}

\usepackage{framed,multirow}

\usepackage{amssymb}
\usepackage{latexsym}

\usepackage{lineno,hyperref}
\usepackage{bm}
\usepackage{amsmath}
\usepackage{multirow}
\usepackage{subfig}

\begin{document}
	

\begin{frontmatter}
	
\title{A versatile lattice Boltzmann model for immiscible ternary fluid flows}

\author[1]{Yuan Yu}

\author[2]{Haihu Liu\corref{cor1}}
\cortext[cor1]{Corresponding author: 
	Tel.: +86 (0) 298 266 5700; } \ead{haihu.liu@mail.xjtu.edu.cn}

\author[1,3]{Dong Liang}

\author[4]{Yonghao Zhang}

\address[1]{School of Engineering, Sun Yat-Sen University, Guangzhou 510006, China}
\address[2]{School of Energy and Power Engineering, Xi'an Jiaotong University, Xi'an 710049, China}
\address[3]{Guangdong Provincial Key Laboratory of Fire Science and Technology, Guangzhou 51006, China}
\address[4]{Department of Mechanical and Aerospace Engineering, University of Strathclyde, Glasgow G1 1XJ, United Kingdom}

\begin{abstract}
We propose a lattice Boltzmann color-gradient model for immiscible ternary fluid flows, which is applicable to the fluids with a full range of interfacial tensions, especially in near-critical and critical states. An interfacial force for N-phase systems is derived based on the previously developed perturbation operator and is then introduced into the model using a body force scheme, which helps reduce spurious velocities. A generalized recoloring algorithm is applied to produce phase segregation and ensure immiscibility of three different fluids, where a novel form of segregation parameters is proposed by considering the existence of Neumann's triangle and the effect of equilibrium contact angle in three-phase junction. The proposed model is first validated with three typical examples, namely the interface capturing for two separate static droplets, the Young-Laplace test for a compound droplet, and the spreading of a droplet between two stratified fluids. This model is then used to study the structure and stability of double droplets in a static matrix. Consistent with the theoretical stability diagram, seven possible equilibrium morphologies are successfully reproduced by adjusting two ratios of the interfacial tensions. By simulating Janus droplets in various geometric configurations, the model is shown to be accurate when three interfacial tensions satisfy a Neumann's triangle. In addition, we also simulate the near-critical and critical states of double droplets where the outcomes are very sensitive to the model accuracy. Our results show that the present model is advantageous to three-phase flow simulations, and allows for accurate simulation of near-critical and critical states.
\end{abstract}

	
\end{frontmatter}



\section{Introduction}
An emulsion is a mixture of a dispersed phase as droplets in another immiscible fluid that forms a continuous phase. Two basic types of emulsions are the oil-in-water (O/W) and water-in-oil (W/O) emulsions\cite{utada2005monodisperse}. Recently, more complex systems referred to as double emulsions and Janus emulsions have received a rapidly growing interest due to their unique properties\cite{bhatia2013review,lamba2015double,chong2015advances} and potential applications\cite{augustin2009nano-,patravale2008novel,ekanem2015structured,ahmed2006enhancement,chen2014microshells:,funakoshi2006lipid}. Double emulsions, also known as `emulsion of emulsion' or `emulsion within emulsion', are emulsions with smaller droplets encapsulated in larger droplets. The shell fluid can serve as a barrier between the core droplets and the outer environment, which makes double emulsions highly desirable for applications in controlled release, separation, and encapsulation\cite{utada2005monodisperse,bhatia2013review,lamba2015double,chong2015advances}. Janus emulsions, which are named after the two-faced Roman god Janus, are highly structured fluids consisting of emulsion droplets that have two distinct physical properties\cite{De1992Soft}. Because of their natural asymmetric ability in the compositions and the shapes, Janus emulsions are often used in the fields that need asymmetry in the shape and the materials. In the applications of emulsions, morphology is one of the most important properties and closely related to other emulsion properties such as rheology, droplet size, relative stability, electrical conductivity and zeta potential\cite{bhatia2013review,lamba2015double,chong2015advances,Vladisavljevi2017Microfluidic}. A number of theoretical and experimental studies have been devoted to identifying different equilibrium morphologies and their transformation. For example, Torza and Mason\cite{torza1969coalescence} studied the droplet morphology in terms of spreading coefficients and obtained the theoretical relationship between the droplet morphology and spreading coefficients. They experimentally observed three equilibrium morphologies of double droplets, i.e. complete engulfing, partial engulfing and non-engulfing, which correspond to three sets of spreading coefficients. Beyond these three equilibrium states, Pannacci et al.\cite{pannacci2008equilibrium} identified several new morphologies of double droplets, and found the non-equilibrium morphologies can have long lifetimes controlled by hydrodynamics, which facilitates the use of double droplets to produce encapsulated particles at early times and Janus particles at longer times. Guzowski et al.\cite{guzowski2012the} presented a detailed theoretical analysis on the possible equilibrium morphologies of double droplets and designed the structure of double emulsions by tuning the volumes of the constituent segments experimentally. As a supplement to theoretical and experimental studies, numerical modelling and simulations are becoming increasingly popular in investigation of the behavior of Janus/double emulsions, which are typical of three-phase flow problems. 

Traditionally, three-phase flows are simulated by solving the macroscopic Navier-Stokes equations together with various approaches to capturing or tracking the interfaces between fluids. Among these approaches, the front-tracking\cite{muradoglu2010front}, volume-of-fluid (VOF)\cite{bonhomme2012inertial}, level-set\cite{zhao1996variational,smith2002projection,saye2011voronoi} and phase-field\cite{boyer2006study,boyer2010cahn-hilliard/navier-stokes,kim2005phase,kim2007phase,kim2009a,kim2012phase-field} methods are commonly used. However, the front-tracking method is not suitable for simulating interface breakup and coalescence; the VOF and level-set methods require either sophisticated interface reconstruction algorithms or unphysical re-initialization processes to represent the interfaces; and the phase-field method yields an interface thickness far greater than its actual value, which may lead to unphysical dissolution of small droplets and mobility-dependent numerical results\cite{liu2018hybrid}. It still remains an open question for the phase-field method to choose an optimal mobility, even for a two-phase flow problem\cite{liu2012three-dimensional}. 

In the past decades, the lattice Boltzmann (LB) method has developed into a promising alternative to the traditional Navier-Stokes-based solvers, for simulating complex flow problems. It is a pseudo-molecular method tracking evolution of the distribution function of an assembly of molecules, built upon microscopic models and mesoscopic kinetic equations\cite{he1997priori}. The LB method has several advantages over the traditional Navier-Stokes-based solvers, e.g. the algorithm simplicity and parallelizability, and the ease of handling complex boundaries\cite{chen1998lattice}. In addition, its kinetic nature allows a simple incorporation of microscopic physics without suffering from the limitations in terms of length and time scales typical of molecular dynamics simulations\cite{aidun2010lattice}. Thus, the LB method is particularly useful in the simulation of multiphase flows. The existing LB models for multiphase flows can be generally classified into four categories: color-gradient model\cite{gunstensen1991lattice,reis2007lattice,liu2012three-dimensional}, interparticle-potential model\cite{shan1993lattice,shan1994simulation,sbragaglia2007generalized,sbragaglia2012emergence,dollet2015two}, phase-field-based model\cite{swift1996lattice,wang2015multiphase,wang2015an}, and mean-field theory model\cite{he1999a}. These models have shown great success as in dealing with two-phase flow problems, and all of them except the mean-field theory model have been extended to the modeling of immiscible ternary fluids, see, e.g. Refs\cite{chen2000a,Shardt2014Simulations,liang2016lattice,semprebon2016ternary,wohrwag2017ternary,shi2016simulation,spencer2010lattice,leclaire2013progress,fu2016numerical}. The ternary color-gradient models\cite{spencer2010lattice,leclaire2013progress} inherit a series of advantages of its two-phase counterpart, such as strict mass conservation for each fluid, flexibly tunable interfacial tensions, and the stability for a broad range of viscosity ratios, and they are well suited to exploring the dynamic processes occurring in ternary fluid systems as previously demonstrated by Fu et al.\cite{fu2016numerical} and Jiang et al.\cite{jiang2017estimation}. The existing color-gradient models, however, commonly suffer from a problem, i.e. three interfacial tensions should satisfy a Neumann's triangle. In industrial processes, surfactants are often added to emulsions to stabilize them against droplet coalescence. The presence of surfactants could significantly modify the interfacial tensions so that the interfacial tensions do not always yield a Neumann's triangle. To correctly predict the dynamical behavior of emulsions, thereby allowing precise control over the droplet geometry and composition, it is necessary for a numerical model to be capable of simulating ternary fluids with a full range of interfacial tensions. On the other hand, it is challenging to simulate the near-critical and critical states of a ternary fluid system where the largest interfacial tension is close to the sum of the other two, as the outcomes are very sensitive to the model accuracy.

In this paper, we develop a LB color-gradient model for simulating immiscible ternary fluids with a full range of interfacial tensions. Based on the perturbation operator developed by Leclaire et al.\cite{leclaire2013progress}, an interfacial force formulation is derived to describe the interactions among different fluids and is then introduced into the model using a body force scheme, which is found to effectively reduce spurious velocities. In addition, the recoloring algorithm proposed by Spencer et al.\cite{spencer2010lattice} is applied to maintain the interfaces and ensure immiscibility of three different fluids, where a new form of segregation parameters is proposed by considering both the existence of Neumann's triangle and the effect of equilibrium contact angle in three-phase junction. The capability and accuracy of this model are first assessed by simulating the interface capturing for two separate static droplets, the Young-Laplace test for a compound droplet, and the spreading of a droplet between two stratified fluids. It is then used to study the structure and stability of double droplets in a static matrix fluid, where we emphasize the model's capability for simulating ternary fluid flows in near-critical and critical states.

\section{Numerical method}
The two-phase color-gradient LB model of Liu et al.\cite{liu2012three-dimensional,liu2015modelling} is extended to the simulation of immiscible ternary fluids. The ternary color-gradient model consists of three steps, i.e. the collision step, the recoloring step and the streaming step. In the collision step, an interfacial force that describes the interactions among different fluids is derived from the perturbation operator presented in Leclaire et al.\cite{leclaire2013progress}, and is then introduced by the body force scheme of Guo et al.\cite{guo2002discrete} In the recoloring step, a novel form of segregation parameters is proposed to ensure accurate phase segregation in three-phase junction and allow for the states where three interfacial tensions between the fluids cannot form a triangle, known as the Neumann's triangle. The distribution functions $f_{i,r}$, $f_{i,g}$ and $f_{i,b}$ are introduced to represent three immiscible fluids, i.e. red fluid, green fluid and blue fluid, where the subscript $i$ is the lattice velocity direction and ranges from 0 to ($n$-1) for a given $m$-dimensional D$m$Q$n$ lattice model. The total distribution function is defined as $f_{i}=\sum_{k}f_{i,k}$ ($k=r$, $g$ or $b$), which undergoes a collision step as
\begin{equation} \label{eq1}
f_{i}^{\dag}\left(\bm{x},t\right)=f_{i}\left(\bm{x},t\right)+\Omega_{i}\left(\bm{x},t\right)+\Phi_i\left(\bm{x},t\right),
\end{equation}
where $f_i \left( {{\bm{x}},t} \right)$ is the total distribution function in the $i$-th velocity direction at the position $\bm{x}$ and the time $t$, $f_i^\dagger $ is the post-collision distribution function, $\Omega _i$ is the Bhatnagar-Gross-Krook (BGK) collision operator, and $\Phi_i$ is the forcing term (also known as perturbation operator), which contributes to the mixed interfacial regions and creates the interfacial tensions between different fluids.

In the BGK collision operator, the total distribution functions are relaxed toward a local equilibrium with a single relaxation time:
\begin{equation}\label{eq2}
\Omega_i({\bm{x}},t)=-\frac{1}{\tau_{f}}\left[f_{i}(\bm{x},t)-f_{i}^{eq}(\bm{x},t)\right],
\end{equation}
where $\tau_f$ is the dimensionless relaxation time, and $f_i^{eq}$ is the equilibrium distribution function of $f_i$.
The equilibrium distribution function is obtained by a second order Taylor expansion of Maxwell-Boltzmann distribution with respect to the local fluid velocity ${\bm{u}}$:
\begin{eqnarray}\label{eq3}
f_{i}^{eq}=w_{i}\rho\left[1+\frac{\bm{e}_{i}\cdot\bm{u}}{c_{s}^{2}}+\frac{\left(\bm{e}_{i}\cdot\bm{u}\right)^{2}}{2c_{s}^{4}}-\frac{\bm{u}\cdot\bm{u}}{2c_{s}^{2}}\right],
\end{eqnarray}
where $\rho=\sum_k \rho_k$ is the total density and $\rho_k$ is the density of the fluid $k$; $c_s$ is
the speed of sound; ${\bm{e}}_i$ is the lattice velocity in the $i$-th direction; and $w_i$ is the weighting factor. For the two-dimensional nine-velocity (D2Q9) model, $\bm{e}_{i}$ is defined as $\bm{e}_0=(0,0)$, $\bm{e}_{1,3}=(\pm c,0)$, $\bm{e}_{2,4}=(0,\pm c)$, $\bm{e}_{5,7}=(\pm c,\pm c)$, and $\bm{e}_{6,8}=(\mp c,\pm c)$, where $c=\delta_x/\delta_t=\sqrt{3}c_s$ with $\delta_x$ and $\delta_t$ being the lattice spacing and time step, respectively (for the sake of simplicity, $\delta_x=\delta_t=1$ is used hereafter); $w_i$ is given by $w_0 = 4/9$, $w_{1-4} = 1/9$ and $w_{5-8} = 1/36$.

Using the concept of a continuum surface force to model the interfacial tension along with the constraints of mass conservation and momentum conservation, Liu et al.\cite{liu2012three-dimensional} derived a generalized expression for the perturbation operator in two-phase simulations. This perturbation operator was later improved by Leclaire et al.\cite{leclaire2013progress} to model the interfacial tensions between different fluids in three-phase simulations. Following Leclaire et al.\cite{leclaire2013progress}, the perturbation operator is given by
\begin{eqnarray}
\label{eq_phi1}
\Phi_i &=& \sum_k \Phi_{i,k}, \\
\Phi_{i,k} &=& \sum_{l,l\neq{k}}\frac{A_{kl}C_{kl}}{2}\left|\bm{G}_{kl}\right|\left[w_{i}\frac{\left(\bm{e}_{i}\cdot\bm{G}_{kl}\right)^{2}}{\left|\bm{G}_{kl}\right|^{2}}-B_{i}\right],
\label{eq_phi_old}
\end{eqnarray}
where $\bm{G}_{kl}=\frac{\rho_{l}}{\rho}\bm{\nabla}\frac{\rho_{k}}{\rho}-\frac{\rho_{k}}{\rho}\bm{\nabla}\frac{\rho_{l}}{\rho}$ is the color gradient~\cite{leclaire2013progress} and is introduced to identify the location of the $k$-$l$ interface, i.e. the interface between the fluid $k$ and the fluid $l$. $C_{kl}$ is a concentration factor that controls the activation of the interfacial tension at the $k$-$l$ interface, and is given by \cite{leclaire2013progress}
\begin{eqnarray}\label{eq15}
C_{kl}=\min\left(10^{6}\frac{\rho_{k}\rho_{l}}{\rho_{k}^{0}\rho_{l}^{0}},1\right),
\end{eqnarray}
where $\rho_{k}^{0}$ is the density of the pure fluid $k$, and $A_{kl}$ is a parameter related to the interfacial tension between the fluids $k$ and $l$, i.e. $\sigma_{kl}=\frac{1}{9}\left(A_{kl}+A_{lk}\right)\tau_f$. The generalized expression for $B_i$ was given by Liu et al.\cite{liu2012three-dimensional} and it was in particular taken as $B_0=-4/27$, $B_{1-4}=2/27$ and $B_{5-8}=5/108$ in the work of Leclaire et al.\cite{leclaire2013progress}. It is worth noting that Eq.(\ref{eq_phi_old}) is not limited to the case with ternary fluids, and can be also applicable to $N$-phase ($N>3$) systems.

Using the Chapman-Enskog multiscale analysis, it is shown that the perturbation operator, given by Eqs.(\ref{eq_phi1}) and (\ref{eq_phi_old}), can lead to the following interfacial force:
\begin{eqnarray}\label{eq11}
\bm{F}_s=-\nabla\cdot\left(\tau_{f}\delta_{t}\sum_{i}{\Phi_{i}\bm{e}_{i}\bm{e}_{i}}\right)
=\sum_k\sum_{l,l\neq k}\nabla\cdot\left[\frac{\sigma_{kl}C_{kl}}{2}\left|\bm{G}_{kl}\right|\left(\bm{I}-\bm{n}_{kl}\bm{n}_{kl}\right)\right],
\end{eqnarray}
where $\bm{n}_{kl}$ is the unit normal vector of the $k$-$l$ interface and is defined by $\bm{n}_{kl}=\bm{G}_{kl}/\left|\bm{G}_{kl}\right|$.

Instead of using Eqs.(\ref{eq_phi1}) and (\ref{eq_phi_old}), the effect of interfacial tension is realized through the body force scheme of Guo et al.\cite{guo2002discrete}, which is able to reduce effectively spurious velocities while keeping high numerical accuracy \cite{liu2015modelling,halliday2006improved}. According to Guo et al.\cite{guo2002discrete}, the forcing term $\Phi_{i}$ in Eq.~(\ref{eq1}) is written as
\begin{eqnarray}\label{eq5}
\Phi_{i}\left(\bm{x},t\right)=w_{i}\left(1-\frac{1}{2\tau_{f}}\right)\left(\frac{\bm{e}_{i}-\bm{u}}{c_{s}^{2}}+\frac{\bm{e}_{i}\cdot\bm{u}}{c_{s}^{4}}\bm{e}_{i}\right)\cdot\bm{F}_s\left(\bm{x},t\right)\delta_{t},
\end{eqnarray}
where the local fluid velocity is defined by the averaged momentum before and after the collision, i.e.,
\begin{equation}
\label{eq_uk1}
\rho{\bm{u}}({\bm{x}},t)=\sum_i f_i({\bm{x}},t){\bm{e}}_i+\frac{1}{2}{\bm{F}}_s({\bm{x}},t)\delta_t.
\end{equation}

In this work, we assume equal densities for the red, green and blue fluids. To allow for unequal viscosities of the three fluids, we determine the local kinematic viscosity $\nu$ by a harmonic mean
\begin{equation}\label{eq_vis_1}
\frac{\rho}{\nu} = \sum_k \frac{\rho_k}{\nu_k},
\end{equation}
where $\nu_k$ ($k=R$, $G$ or $B$) is the kinematic viscosity of the fluid $k$.  The local relaxation time $\tau_f$ can be calculated from the local viscosity using the following equation:
\begin{equation}\label{eq_tau}
\nu = \left(\tau_f-\frac{1}{2}\right)c_s^2\delta_t.
\end{equation}

The partial derivatives in the interfacial force ${\bm{F}}_s$ should be evaluated through suitable difference schemes. To minimize the discretization errors, the fourth-order isotropic finite difference scheme
\begin{equation}\label{eq_fd1}
\partial_{\alpha} \varphi\left( {\bf{x}}, t\right) =\frac{1}{c_s^2}\sum_i w_i \varphi \left( {\bf{x}}+{\bf{e}}_i\delta_t, t\right)e_{i\alpha},
\end{equation}
is used to evaluate the derivatives of a variable $\varphi$.

Although the forcing term generates the interfacial tensions, it does not guarantee the immiscibility of different fluids. In order to minimize the mixing of the fluids, a recoloring step is applied. Based on the pioneering work of D'Ortona et al.\cite{d1995two}, Latva-Kokko and Rothman\cite{latva2005diffusion} developed a recoloring algorithm to demix two immiscible fluids, which can overcome the lattice pinning problem and creates a symmetric distribution of particles around the interface so that unphysical spurious velocities can be effectively reduced. This recoloring algorithm was later generalized by Spencer et al.\cite{spencer2010lattice} to three-phase fluid flows. Following Spencer et al.\cite{spencer2010lattice}, the recolored distribution functions of the fluid $k$ ($k=r$, $g$ or $b$) are
\begin{eqnarray}\label{eq6}
f_{i,k}^{\ddagger}\left(\bm{x},t\right)=\frac{\rho_{k}}{\rho}f_{i}^{\dagger}\left(\bm{x},t\right)+\sum\limits_{l,l\neq{k}}\beta_{kl}w_{i}\frac{\rho_{k}\rho_{l}}{\rho}\bm{n}_{kl}\cdot \bm{e}_{i},
\end{eqnarray}
where $f_{i,k}^{\ddagger}$ is the recolored distribution functions of the fluid $k$, and $\beta_{kl}$ is a segregation parameter related to the thickness of the $k$-$l$ interface. It should be noted that $\beta_{kl}=\beta_{lk}$ in order to conserve mass and momentum during the recoloring process.

\begin{figure}[!htb]
	\centering
	\includegraphics[scale=0.35]{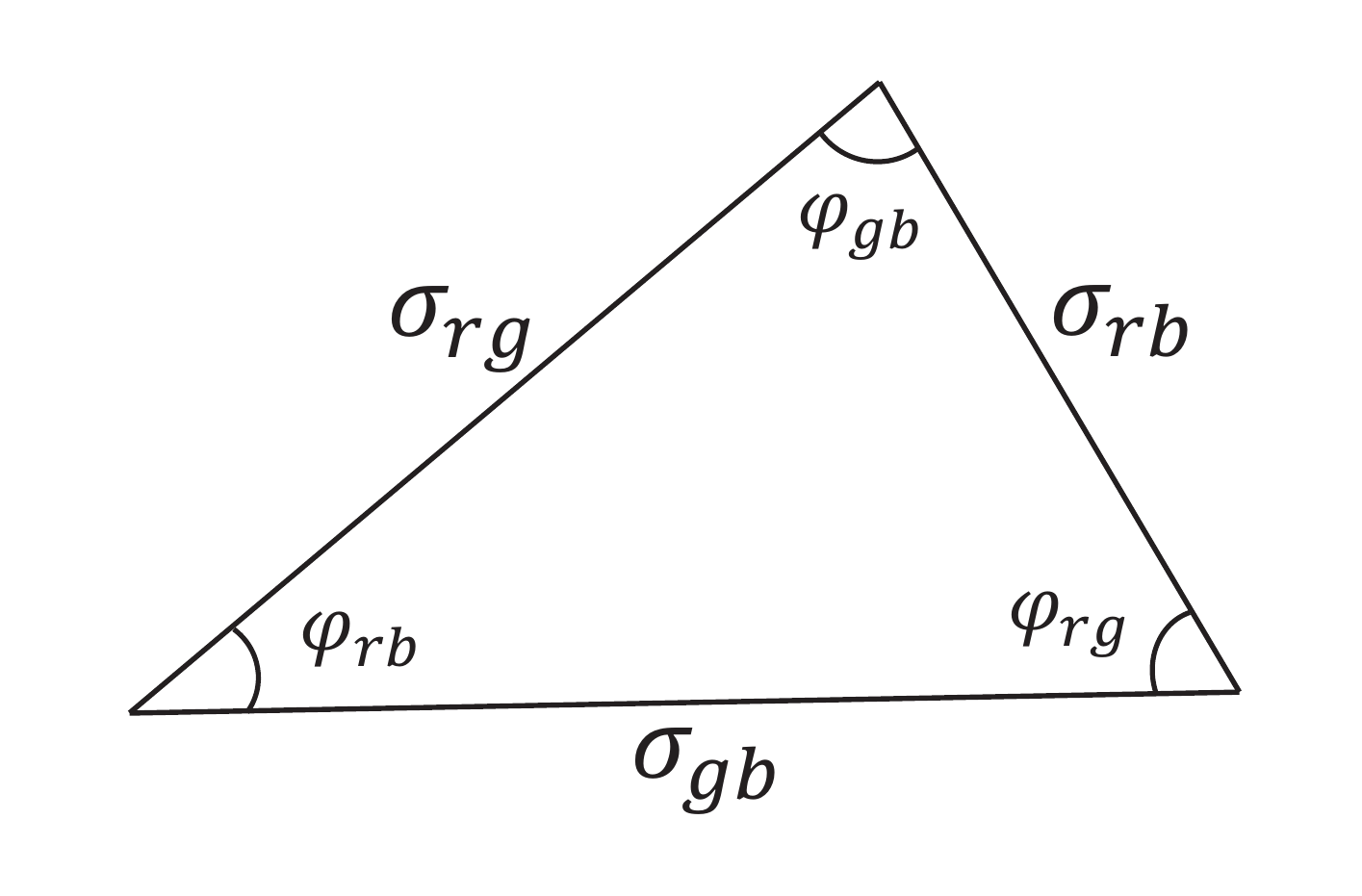}
	\caption{\ Neumann's triangle}
	\label{neumann}
\end{figure}

For the ternary fluids and when three interfacial tensions satisfy a Neumann's triangle (see Fig.~\ref{neumann}), the equilibrium contact angle $\varphi_{kl}$ will be formed between the fluids in three-phase junction, and it is related to the interfacial tensions by
\begin{eqnarray}\label{eq_theta_kl}
\cos(\varphi_{kl})=\frac{\sigma_{mk}^{2}+\sigma_{ml}^{2}-\sigma_{kl}^{2}}{2\sigma_{mk}\sigma_{ml}}.
\end{eqnarray}
Spencer et al.~\cite{spencer2010lattice} theoretically showed that in three-phase junction, there should be a relationship between $\varphi_{kl}$ and the (relative) interface thickness, which is controlled by the segregation parameter $\beta_{kl}$. Hence, it is of great importance to select a proper $\beta_{kl}$ in three-phase simulations. Several different forms of $\beta_{kl}$ have been provided in literature. Spencer et al. \cite{spencer2010lattice} proposed the first expression for the segregation parameters, which is given by
\begin{eqnarray}\label{eq16}
\left\{
\begin{aligned}
\beta_{rg}=&\beta^{0} \\
\beta_{rb}=&\beta^{0}\left[1+\frac{27\rho_{r}\rho_{g}\rho_{b}}{\rho^{3}}\left(\sin{\varphi_{gb}}-1\right)\right] \\
\beta_{gb}=&\beta^{0}\left[1+\frac{27\rho_{r}\rho_{g}\rho_{b}}{\rho^{3}}\left(\sin{\varphi_{rb}}-1\right)\right]
\end{aligned}
\right.,
\end{eqnarray}
where $\beta^{0}$ is the reference segregation parameter. Clearly, the segregation parameters in Eq.~(\ref{eq16}) will degenerate into $\beta_{kl}=\beta^{0}$ at an interface where only two fluids are present. So it is suggested to take $\beta^{0}=0.7$ to be consistent with the segregation parameter in the two-phase color-gradient model\cite{liu2012three-dimensional}. Leclaire et al. \cite{leclaire2013progress} improved the segregation parameters of Spencer et al. \cite{spencer2010lattice} by setting $\beta_{kl}=\beta^0$  for the largest $\varphi_{kl}$ in the Neumann's triangle,
\begin{eqnarray}\label{eq17}
\beta_{kl}=\left\{
\begin{aligned}
\beta^{0}& &\text{$kl$ with $\varphi_{max}$}  \\
\beta^{0}&+\beta^{0}C_{t}\left[\sin{\left(\pi-\varphi_{max}-\varphi_{kl}\right)}-1\right] &\text{otherwise}
\end{aligned}
\right.,
\end{eqnarray}
where $\varphi_{max}=\max(\varphi_{kl})$ and $C_{t}=\min\left(\frac{35\rho_{r}\rho_{g}\rho_{b}}{\rho^{3}},1\right)$. Leclaire et al.\cite{leclaire2013progress} also mentioned to use $\beta_{kl}=\beta^{0}$ when the Neumann's triangle does not exist. Clearly, Eq.~(\ref{eq17}) will degradate to Eq.~(\ref{eq16}) when $\varphi_{max}=\varphi_{rg}$. Althogh Eqs.(\ref{eq16}) and (\ref{eq17}) work to some extent especially when the Neumann's triangle exists, they cannot accurately simulate the critical state where the largest interfacial tension equals the sum of the other two, which will be shown later. Recently, Fu et al.\cite{fu2016numerical} seemed to have also noticed that Eqs.(\ref{eq16}) and (\ref{eq17}) do not always produce convincing results in three-phase simulations, so they simply selected a constant $\beta_{kl}$, i.e.

\begin{equation}\label{eq_fu}
\beta_{kl}=\beta^{0}.
\end{equation}
It is evident that the dependence of $\beta_{kl}$ on $\varphi_{kl}$ is not considered in Eq.(\ref{eq_fu}), and thus incorrect results may be obtained, e.g. in the critical state.

To overcome the aforementioned drawbacks associated with the existing $\beta_{kl}$, a novel form of segregation parameters is proposed. First, we determine whether the Neumann's triangle exists by calculating
\begin{eqnarray}\label{eq18}
X_{kl}=\frac{\sigma_{mk}^{2}+\sigma_{ml}^{2}-\sigma_{kl}^{2}}{2\sigma_{mk}\sigma_{ml}}.
\end{eqnarray}
It is easily seen from Eq.(\ref{eq_theta_kl}) that the Neumann's triangle will exist if $\left|X_{kl}\right|<1$ for all $kl$. Then, the segregation parameter $\beta_{kl}$ is defined as a continuous function of $X_{kl}$:
\begin{eqnarray}\label{eq19}
\beta_{kl}=\beta^{0}+\beta^{0}\min\left(\frac{35\rho_{r}\rho_{g}\rho_{b}}{\rho^{3}},1\right)g\left(X_{kl}\right),
\end{eqnarray}
where
\begin{eqnarray}\label{eq20}
g\left(X_{kl}\right)=\left\{
\begin{aligned}
&1 & X_{kl}<-1\\
&1-\sin{\left(\arccos\left(X_{kl}\right)\right)}  & -1\leq X_{kl}< 0 \\
&\sin{\left(\arccos\left(X_{kl}\right)\right)}-1  & 0\leq X_{kl}\leq 1  \\
&-1 & 1<X_{kl}
\end{aligned}
\right..
\end{eqnarray}

It should be noted in three-phase junction that Eqs.(\ref{eq19}) and (\ref{eq20}) are derived based on the following relationship:

\begin{equation}\label{eq_sin}
\frac{\beta_{rg}}{\sin(\varphi_{rg})}=\frac{\beta_{rb}}{\sin(\varphi_{rb})}=\frac{\beta_{gb}}{\sin(\varphi_{gb})},
\end{equation}
which is consistent with the nature of diffuse interfaces, thus leading to more accurate results than using other forms of $\beta_{kl}$. Moreover, the proposed $\beta_{kl}$ works well no matter if the Neumann's triangle exists or not.

After the recoloring step, the red, green and blue distribution functions propagate to the neighboring lattice nodes, known as the propagation or streaming step:

\begin{equation}\label{30}
f_{i,k}\left(\bm{x}+\bm{e}_{i}\delta_{t},t+\delta_{t}\right)=f_{i,k}^{\ddagger}\left(\bm{x},t\right), \quad k=\{r,g,b\}
\end{equation}
with the post-propagation distribution functions used to compute the densities of colored fluids by $\rho_k=\sum_i f_{i,k}$.

\section{Numerical Validations}
\subsection{Interface capturing}
\label{sec_interface}
\begin{figure}[!htb]
	\centering
	\includegraphics[scale=0.35]{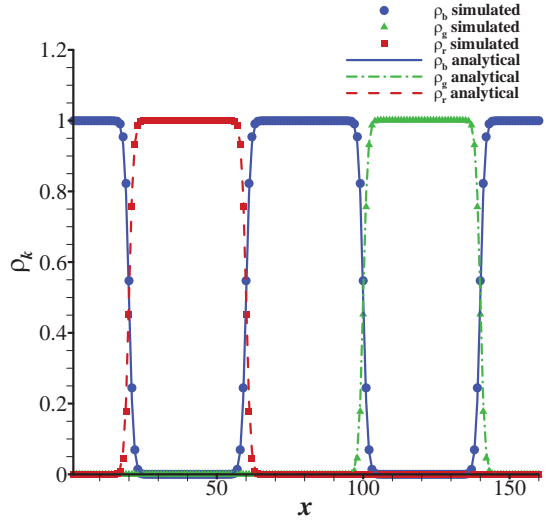}
	\caption{\ The equilibrium density distributions of three different fluids for two separate static droplets immersed in a third fluid.}
	\label{interface}
\end{figure}

We first consider two separate static droplets immersed in another fluid (say blue fluid) to validate the present model for interface capturing. Initially, a red droplet and a green droplet, both having equal radius $R=20$, are placed in a $200\times100$ lattice domain, and their centers are located at $(x_r,y_r)=(50,50)$ and $(x_g,y_g)=(150,50)$, respectively. Considering the distance between two droplets, each droplet interface is essentially a two-phase region, so the equilibrium density distributions at $y=50$ can be analytically expressed as\cite{liu2017a}
\begin{subequations}\label{eq21}
	\begin{eqnarray}\label{eq21:1}
	\frac{\rho_{r}}{\rho}\left(x\right)=0.5+0.5\tanh\left[\frac{R-\sqrt{(x-x_r)^{2}}}{\xi}\right],
	\end{eqnarray}
	\begin{eqnarray}\label{eq21:2}
	\frac{\rho_{g}}{\rho}\left(x\right)=0.5+0.5\tanh\left[\frac{R-\sqrt{(x-x_g)^{2}}}{\xi}\right],
	\end{eqnarray}
	\begin{eqnarray}\label{eq21:3}
	\frac{\rho_{b}}{\rho}\left(x\right)=1-\frac{\rho_{r}}{\rho}\left(x\right)-\frac{\rho_{g}}{\rho}\left(x\right),
	\end{eqnarray}
\end{subequations}
for the red, green and blue fluids, respectively. Here, the parameter $\xi$ is a measure of the interface thickness related to $\beta^{0}$ by $\xi=1/(6k\beta^{0})$~\cite{riaud2014lattice-boltzmann}, and $k$ is a geometric constant that is determined by \cite{liu2017a}
\begin{eqnarray}\label{eq_k}
k=\frac{1}{2}\sum_{i}{\frac{w_{i}\bm{e}_{i}\bm{e}_{i}}{\left|\bm{e}_{i}\right|}}.
\end{eqnarray}
For the D2Q9 model, one can obtain $k\approx0.1504$ from Eq.(\ref{eq_k}), and thus $\xi\approx1.5831$ for $\beta^{0}=0.7$.
The simulation is run with the interfacial tensions $\sigma_{rg}=\sigma_{rb}=\sigma_{gb}=0.01$ and the viscosities $\nu_r=\nu_g=\nu_b=0.1$. Periodic boundary conditions are applied in both the $x$ and $y$ directions. Fig.~\ref{interface} shows the simulated density distributions of the red, green and blue fluids along $y=50$ in the steady state, and the corresponding analytical solutions, given by Eq.(\ref{eq21}), are also shown for comparison. Clearly, the simulated density distributions are all in good agreement with the analytical solutions, indicating that
the present color-gradient LBM can correctly model and capture phase interfaces.

\subsection{Young-Laplace test}
A compound droplet, which consists of an inner droplet encapsulated by another immiscible fluid, suspended in a third fluid, is simulated to assess whether the interfacial tensions are correctly modelled. The computational domain is taken as $160\times160$, and it is filled with three different fluids, which are initialized as
\begin{eqnarray}
\left\{
\begin{aligned}
\rho_r=1, \quad \rho_g=\rho_b=0 & & (x-80)^2+(y-80)^2\le R_r^2  \\
\rho_g=1, \quad \rho_r=\rho_b=0 && \quad R_r^2<(x-80)^2+(y-80)^2\le R_g^2 \\
\rho_b=1, \quad \rho_r=\rho_g=0 && \text{otherwise}
\end{aligned}
\right.
\end{eqnarray}
with $R_g=2R_r$. This gives the initial condition that a compound droplet is located in the center of the computational domain. The interfacial tensions and the fluid viscosities are all kept the same as those used in Section~\ref{sec_interface}, and the periodic boundary conditions are used in both $x$ and $y$ directions. According to the Young-Laplace's law, when the system reaches the equilibrium state, the pressure difference $\Delta p$ across an interface is related to the interfacial tension $\sigma$ by
\begin{eqnarray}\label{eq_laplace}
\Delta p=\frac{\sigma}{R},
\end{eqnarray}
where $R$ is the radius of the interface curvature. Eq.(\ref{eq_laplace}) allows us to quantify the modeling accuracy of interfacial tensions through the relative error
\begin{eqnarray}\label{eq22}
\epsilon=\frac{\left|\Delta{p_{gb}}R_{g}+\Delta{p_{rg}}R_{r}-\left(\sigma_{gb}+\sigma_{rg}\right)\right|}{\sigma_{gb}+\sigma_{rg}}\times 100\%.
\end{eqnarray}

\begin{table*}[!htb]
	\small
	\caption{\ The relative errors of interfacial tensions for various values of $R_r$.}
	\label{tab1}
	\begin{tabular*}{\textwidth}{@{\extracolsep{\fill}}lllll}
		\hline
		$R_{r}$&15&20&25&30\\
		\hline
		$\epsilon$ & 1.3\% & 0.95\% & 0.83\% & 0.57\% \\
		\hline
	\end{tabular*}
\end{table*}

\begin{table*}[!htb]
	\small
	\caption{\ The maximum spurious velocities ($\left|\bm{u}\right|_{max}$) obtained with two different forcing methods for various $R_r$.}
	\label{tab2}
	\begin{tabular*}{\textwidth}{@{\extracolsep{\fill}}llllll}
		\hline
		$R_{r}$& &15&20&25&30\\
		\hline
		\multirow{2}{*}{$\left|\bm{u}\right|_{max}\times{10^{5}}$}
		&Present forcing method     & 1.68& 1.69& 1.70& 1.71 \\
		\cline{2-6}
		&Forcing method of Leclaire et al.  & 3.15& 3.28& 3.27& 3.29 \\
		\hline
	\end{tabular*}
\end{table*}

Table~\ref{tab1} shows the relative errors of interfacial tensions for different values of $R_r$. All the relative errors $\epsilon$ are below $1.5\%$, suggesting that our LBM results are in excellent agreement with the Young-Laplace's law. In addition to the present forcing method, i.e. Eqs.(\ref{eq11}) and (\ref{eq5}), the interfacial tension effects can also be realized by the forcing method of Leclaire et al.\cite{leclaire2013progress}, i.e. Eqs.(\ref{eq_phi1}) and (\ref{eq_phi_old}). It is of interest to compare the effect of these two different forcing methods on spurious velocities. Table~\ref{tab2} shows the maximum spurious velocities ($\left|\bm{u}\right|_{max}$) for various $R_r$, where the values of $\left|\bm{u}\right|_{max}$ are magnified by $10^5$ times. It is seen that the maximum spurious velocities are almost independent of $R_r$ for either forcing method, and that the present spurious velocities are always smaller than those obtained with the forcing method of Leclaire et al.~\cite{leclaire2013progress}.

\subsection{Spreading of a droplet between two stratified fluids}
To assess the overall performance of the proposed model, we simulate the spreading of a droplet between two other immiscible fluids. The computational domain is set to be $160\times 160$ lattices. Initially, a red circular droplet with the radius $R=20$ is placed in the center of the computational domain, and the green and blue fluids are allocated to the lower and upper halves of the computational domain outside the droplet. Periodic boundary conditions are used in both the $x$ and $y$ directions. Depending on the values of the interfacial tensions, two different spreading phenomena can be observed, i.e. partial spreading and complete spreading.

\begin{figure}[!htb]
	\centering
	\includegraphics[scale=0.4]{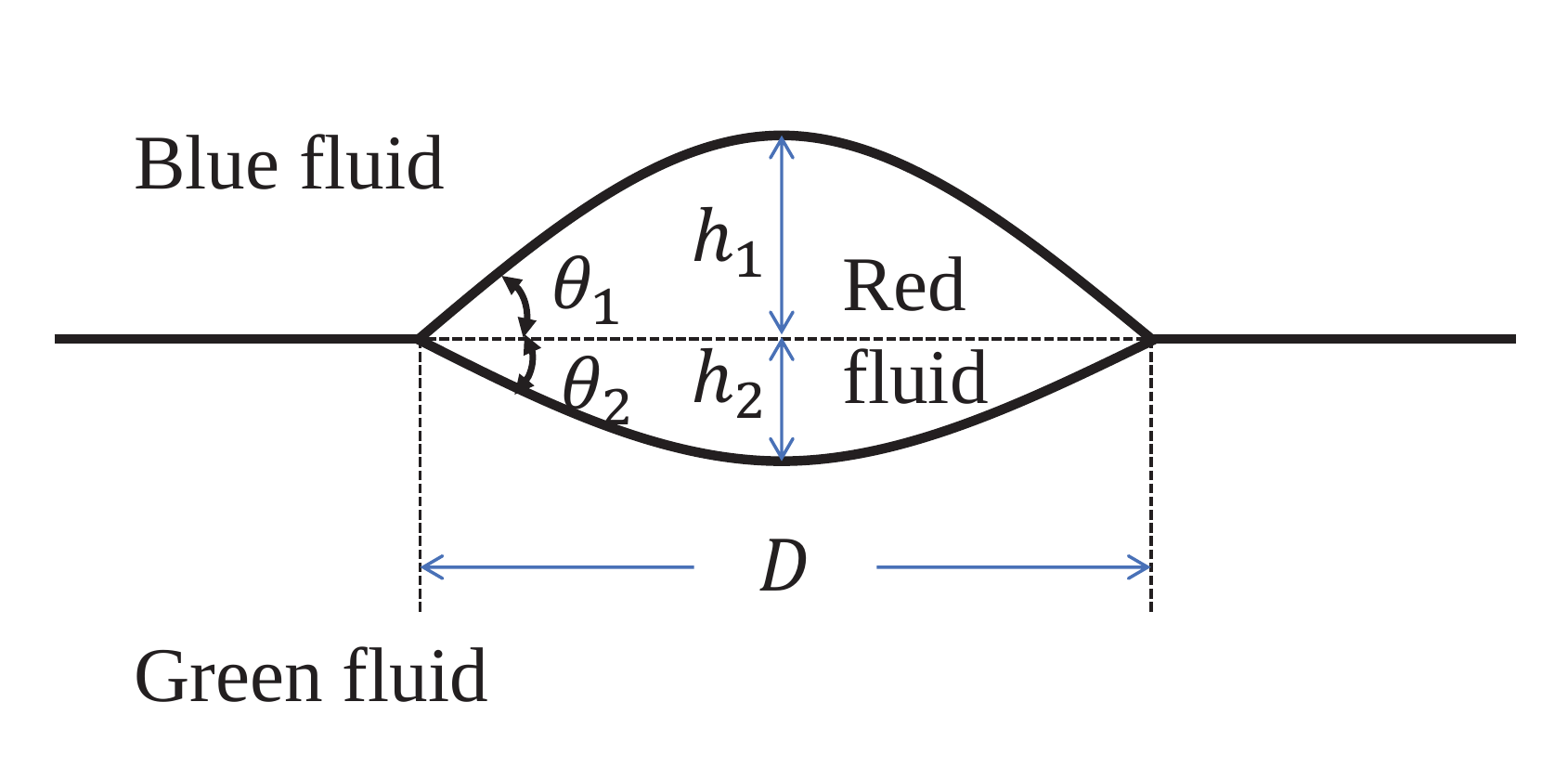}
	\caption{\ The shape of a liquid lens at equilibrium.}
	\label{lens}
\end{figure}

\begin{figure*}
	\centering
	\subfloat[]{\includegraphics[scale=0.3]{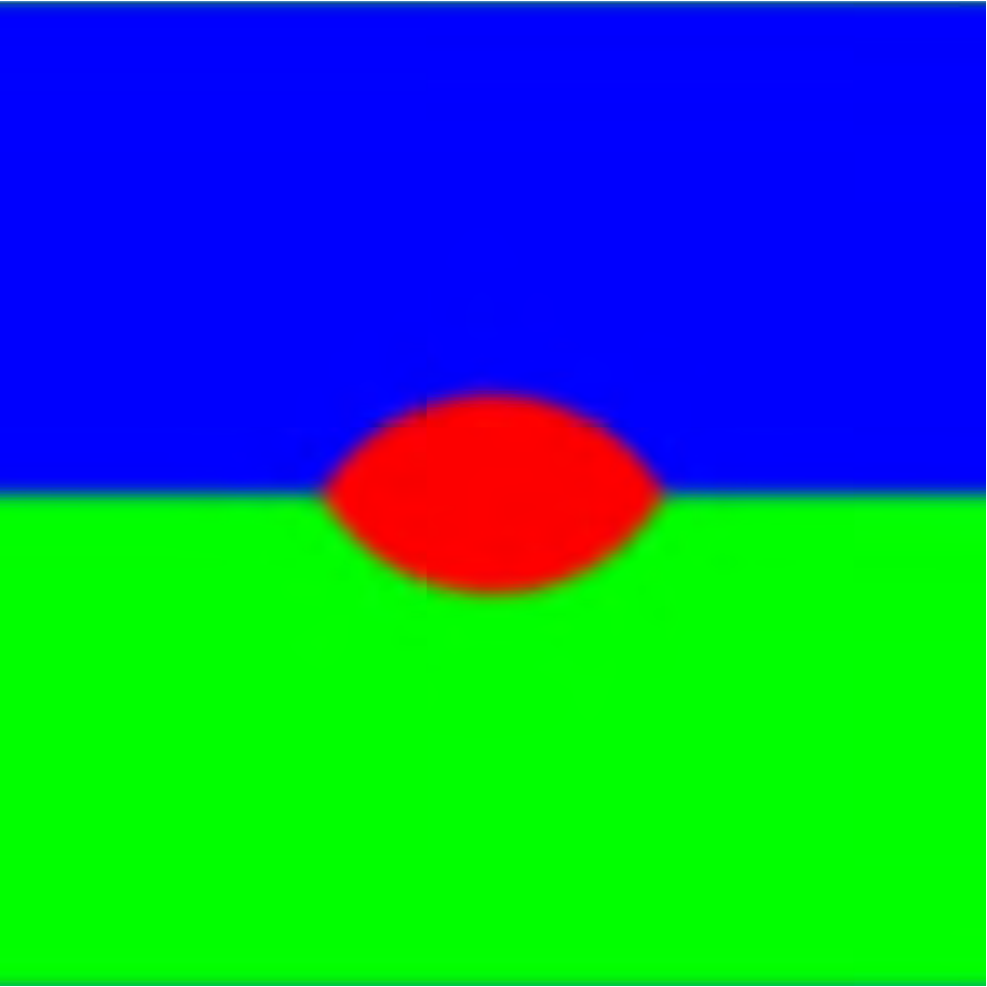}}
	\subfloat[]{\includegraphics[scale=0.3]{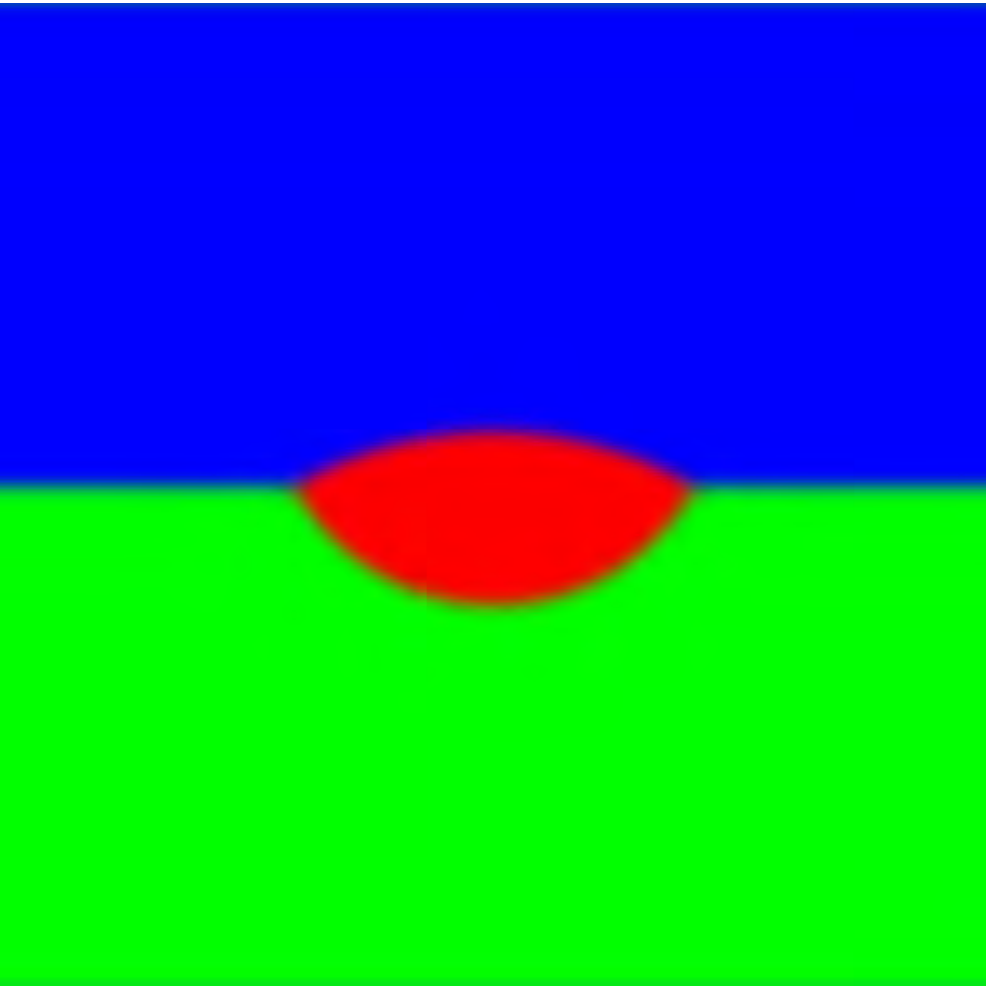}}
	\subfloat[]{\includegraphics[scale=0.3]{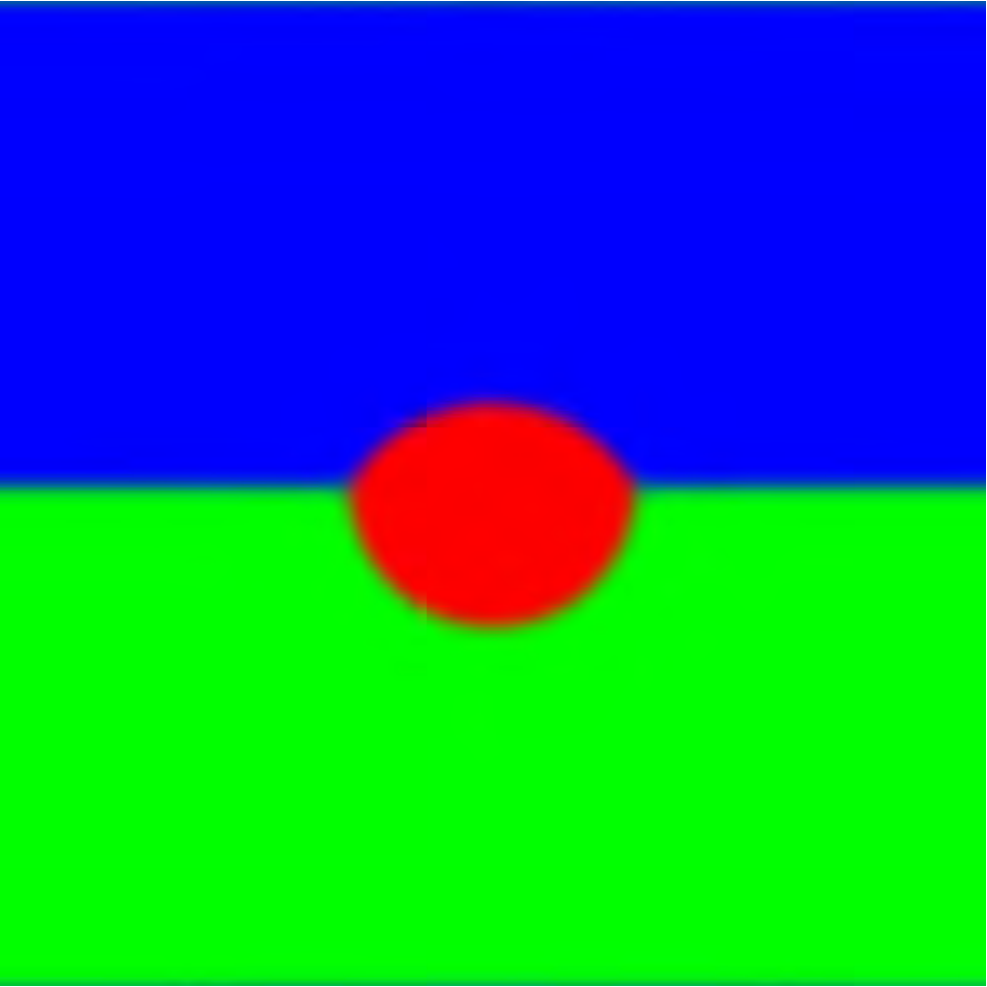}}
	\subfloat[]{\includegraphics[scale=0.3]{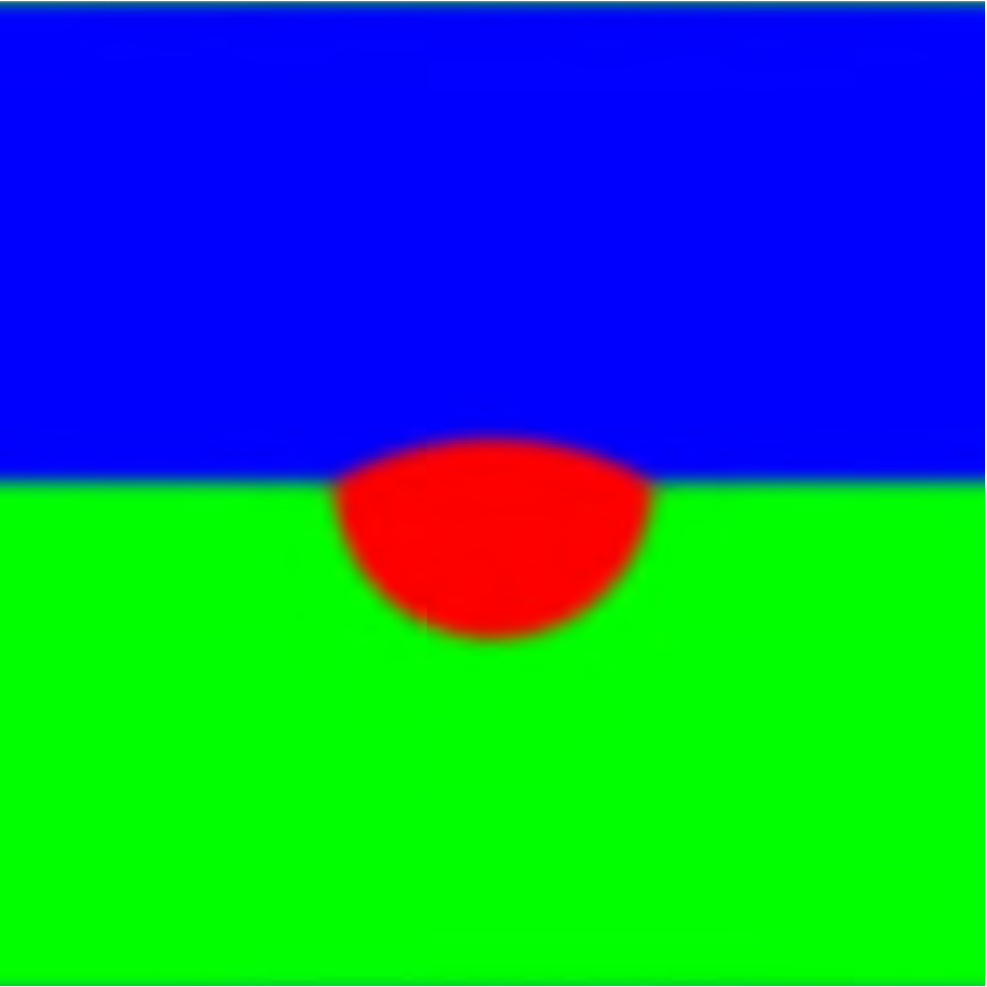}}
	\caption{\ Final fluid distributions in the cases of partial spreading for (a) $\sigma_{rb}=0.01$, $\sigma_{rg}=0.01$; (b) $\sigma_{rb}=0.0087$, $\sigma_{rg}=0.005$; (c) $\sigma_{rb}=0.02$, $\sigma_{rg}=0.0173$; (d) $\sigma_{rb}=0.0115$, $\sigma_{rg}=0.0058$. The third interfacial tension is fixed at $\sigma_{gb}=0.01$.}
	\label{fig:partial_lens}
\end{figure*}

We first consider the partial spreading, where three interfacial tensions yield a Neumann's triangle. In a partial spreading, the droplet can eventually reach a steady lens shape, which is often characterized by the lens length $D$ and the heights $h_1$ and $h_2$ (see Fig.~\ref{lens}). The lens length and heights can be analytically given as \cite{rowlinson2013molecular,jiang2017estimation}
\begin{subequations}\label{eq23}
	\begin{eqnarray}\label{eq23:1}
	D=2\sqrt{\frac{A}{\sum\limits_{i=1}^{2}{\frac{1}{\sin\theta_{i}}\left(\frac{\theta_{i}}{\sin\theta_{i}}-\cos\theta_{i}\right)}}},
	\end{eqnarray}\label{eq23:2}
	\begin{eqnarray}
	h_{i}=\frac{D}{2}\left(\frac{1-\cos\theta_{i}}{\sin\theta_{i}}\right)\quad \text{with}\quad i=1,2,
	\end{eqnarray}
\end{subequations}
where $A$ is the area of the red droplet; $\theta_{1}=\varphi_{rg}$ and $\theta_{2}=\varphi_{rb}$ are the equilibrium contact angles that can be calculated from Eq.(\ref{eq_theta_kl}). Four groups of interfacial tensions are simulated with a constant $\sigma_{gb}$ of $0.01$ but varying $\sigma_{rb}$ and $\sigma_{rg}$, i.e., (a) $\sigma_{rb}=0.01$ and $\sigma_{rg}=0.01$, (b) $\sigma_{rb}=0.0087$ and $\sigma_{rg}=0.005$, (c) $\sigma_{rb}=0.0173$ and $\sigma_{rg}=0.02$, (d) $\sigma_{rb}=0.0058$ and $\sigma_{rg}=0.0115$. The fluid viscosities are all kept at $0.1$, and the final fluid distributions are shown in Fig.~\ref{fig:partial_lens}. As expected, the droplet exhibits a lens shape in each of the cases considered, and the geometrical sizes ($D$, $h_1$ and $h_2$) of the lens are case dependent. Based on the fluid distributions, we also quantify the geometrical sizes of the lens, and compare the simulated results with the analytical predictions from Eq.(\ref{eq23}). It is seen in Table~\ref{tab3} that the simulated results (denoted by $D^{s}$, $h_1^{s}$ and $h_2^{s}$) agree well with the analytical predictions (denoted by $D^{a}$, $h_1^{a}$ and $h_2^{a}$) with the relative errors (defined by $E(\chi)=\frac{|\chi^{a}-\chi^{s}|}{\chi^{a}}\times 100\%$, where $\chi=D$, $h_1$ or $h_2$) all around $1\%$ except in the cases of small contact angles. The increased errors at small contact angles are attributed to the low resolution in sharp corners, which were also found by Jiang and Tsuji\cite{jiang2017estimation}.

\begin{table*}
	\small
	\caption{\ Comparison between the analytical predictions and simulated results for the geometrical sizes of the deformed droplet.}
	\label{tab3}
	\begin{tabular*}{\textwidth}{@{\extracolsep{\fill}}cccccccccccc}
		\hline
		Case &$D^{a}$ &$h_{1}^{a}$ & $h_{2}^{a}$ &$D^{s}$ &$h_{1}^{s}$ & $h_{2}^{s}$ & $E\left(D\right)$ &$E\left(h_{1}\right)$ &$E\left(h_{2}\right)$ \\
		\hline
		(a) &55.34& 15.97& 15.97& 53.59& 15.92& 15.92& 3.17\%& 0.37\%& 0.38\%\\
		(b) &65.18& 8.73& 18.81& 61.79& 8.79& 18.97& 5.21\%& 0.59\%& 0.82\%\\
		(c) &45.84& 13.23& 22.92& 45.39& 13.35& 22.51& 0.99\%& 0.85\%& 1.80\%\\
		(d) &50.99& 6.83& 25.49& 50.93& 6.77& 25.22& 0.13\%& 0.94\%& 1.09\% \\
		\hline
	\end{tabular*}
\end{table*}

\begin{figure*}[!htb]
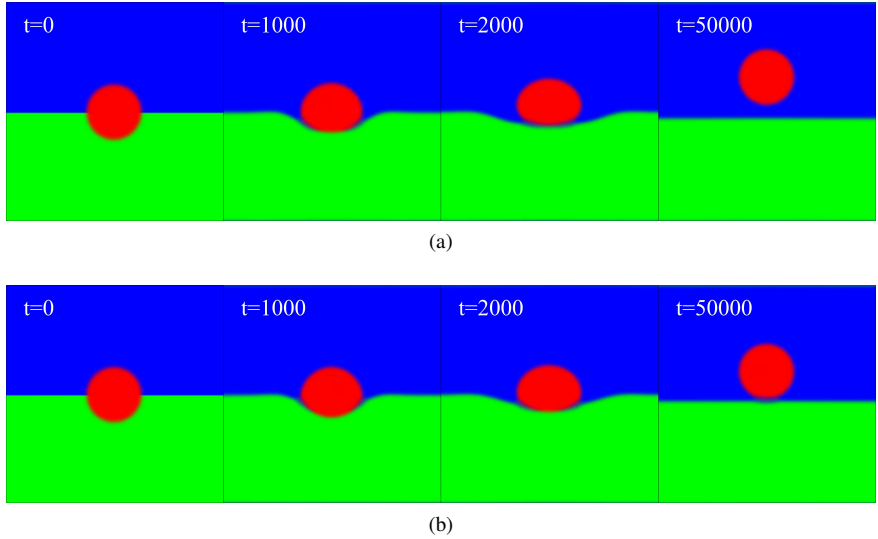

	\centering
	\subfloat[]{\includegraphics[scale=0.04]{Sbt0.pdf}}\\
	\subfloat[]{\includegraphics[scale=0.04]{Seq0.pdf}}
	\caption{\ Time evolution of the interface in the cases of complete spreading for (a) $\sigma_{rg}=0.01$ and (b) $\sigma_{rg}=0.015$. The other two interfacial tensions are fixed at $\sigma_{gb}=\sigma_{rb}=0.005$. Note that the system has reached the steady state at $t=50000$ in each case.}
	\label{fig_lens_S}
\end{figure*}

We then consider the complete spreading, where three interfacial tensions cannot yield a Neumann's triangle. Two different cases of complete spreading are simulated for $\sigma_{rg}=0.01$ and $\sigma_{rg}=0.015$ at $\sigma_{gb}=\sigma_{rb}=0.005$. Clearly, $\sigma_{rg}=\sigma_{gb}+\sigma_{rb}$ in the first case, which corresponds to the critical state; whereas $\sigma_{rg}>\sigma_{gb}+\sigma_{rb}$ in the second case, which corresponds to the supercritical state. Fig.~\ref{fig_lens_S} shows the time evolution of the interface in both cases for a constant fluid viscosity of $0.05$. We can see that in the critical state, the red droplet sits exactly on the $gb$ interface in the end; whereas in the supercritical state, it bounces off the $gb$ interface and rises up to the blue fluid.

\section{Structure and stability of double droplets}
Double emulsions have received considerable attention because of their potential applications in food science, cosmetics, pharmaceuticals and medical diagnostics. Since emulsion properties and functions are related to the droplet geometry and composition, it is of great importance, from a numerical point of view, to accurately predict the topological structure of double droplets when dispersed in another immiscible fluid.

\subsection{Stability diagram for double droplets}
\begin{figure*}[!htb]
	\centering
	\includegraphics[scale=0.5]{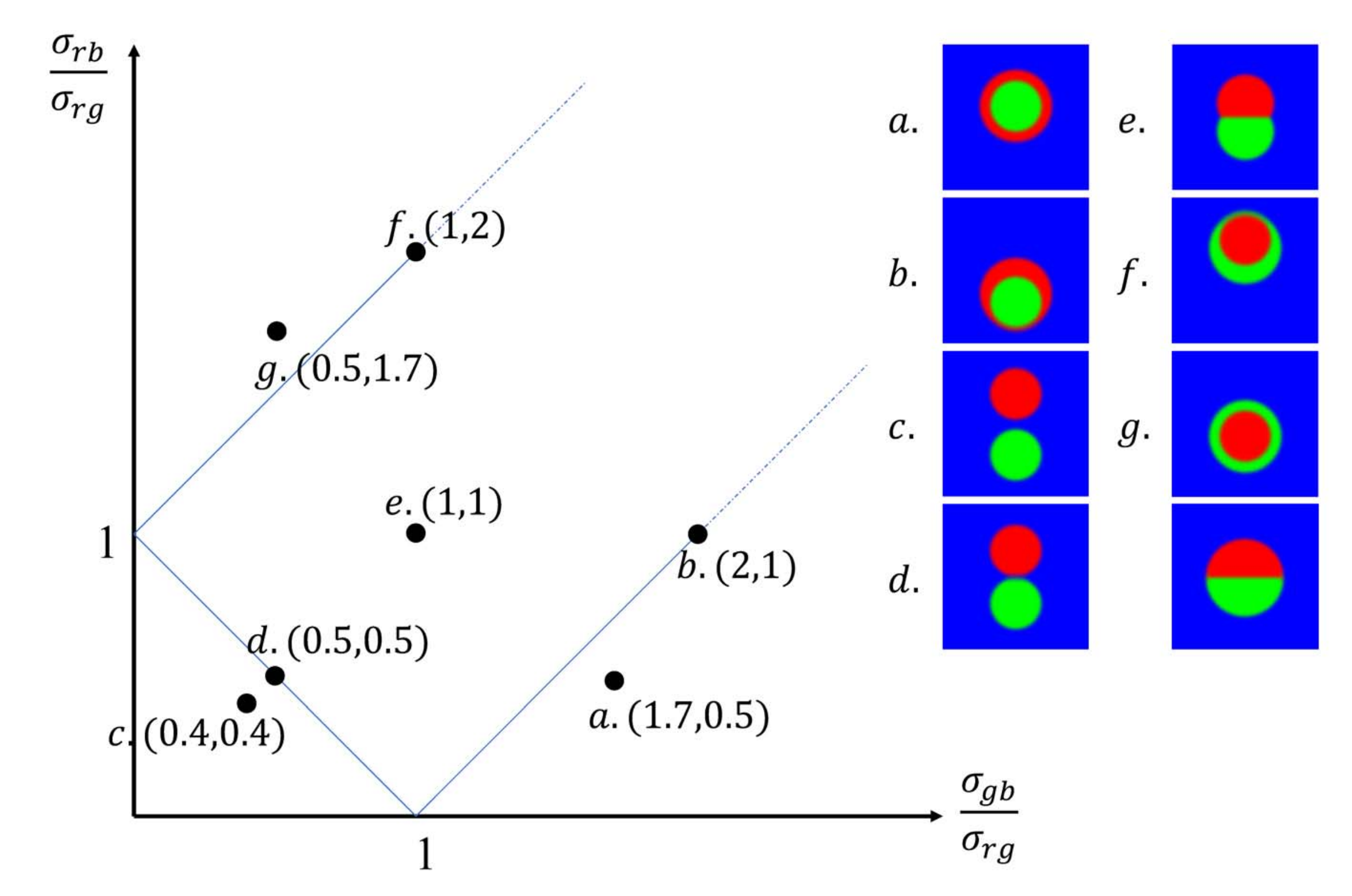}
	\caption{\ Stability diagram representing possible morphologies of double droplets (left panel) and equilibrium shapes of the droplets for the typical cases marked in the stability diagram (right panel). The red lines represent the critical morphologies or the transitions between the regions of complete engulfing, partial engulfing and non-engulfing.}
	\label{double_droplet}
\end{figure*}

Consider a pair of equal-sized droplets, consisting of red and green fluids and initially sitting next to each other, immersed in the third fluid (blue fluid). Based on the theoretical analysis, Guzowski et al.\cite{guzowski2012the} presented a stability diagram that describes the possible topologies of double droplets and their transitions in terms of two ratios of the interfacial tensions (see the left panel of Fig.~\ref{double_droplet}). In the stability diagram, seven typical cases (represented by the solid points) are simulated to examine if the present model is able to reproduce the correct morphologies of double droplets. These typical cases are (i) $\frac{\sigma_{gb}}{\sigma_{rg}}=1.7$ and $\frac{\sigma_{rb}}{\sigma_{rg}}=0.5$, (ii) $\frac{\sigma_{gb}}{\sigma_{rg}}=2$ and $\frac{\sigma_{rb}}{\sigma_{rg}}=1$, (iii) $\frac{\sigma_{gb}}{\sigma_{rg}}=0.4$ and $\frac{\sigma_{rb}}{\sigma_{rg}}=0.4$, (iv) $\frac{\sigma_{gb}}{\sigma_{rg}}=0.5$ and $\frac{\sigma_{rb}}{\sigma_{rg}}=0.5$, (v) $\frac{\sigma_{gb}}{\sigma_{rg}}=1$ and $\frac{\sigma_{rb}}{\sigma_{rg}}=1$, (vi) $\frac{\sigma_{gb}}{\sigma_{rg}}=1$ and $\frac{\sigma_{rb}}{\sigma_{rg}}=2$, and (vii) $\frac{\sigma_{gb}}{\sigma_{rg}}=0.5$ and $\frac{\sigma_{rb}}{\sigma_{rg}}=1.7$, which cover all the possible morphologies identified by Guzowski et al.\cite{guzowski2012the}.

The computational domain is taken to be $[1,120]\times[1,120]$, and the initial fluid distributions are
\begin{eqnarray}
&\rho_{r}(x,y)=0.5+0.5\tanh\left[\frac{R-\sqrt{(x-60.5)^{2}+(y-60.5-R)^{2}}}{\xi}\right], \\
&\rho_{g}(x,y)=0.5+0.5\tanh\left[\frac{R-\sqrt{(x-60.5)^{2}+(y-60.5+R)^{2}}}{\xi}\right], \\
&\rho_{b}(x,y)=1-\rho_{r}(x,y)-\rho_{g}(x,y),
\end{eqnarray}
where the droplet radius $R=20$ lattices. The periodic boundary conditions are used in both the $x$ and $y$ directions. All the fluids are assumed to have equal viscosity of $0.1$, and the interfacial tension $\sigma_{rg}$ is fixed at $0.01$. The simulations are run until an equilibrium state is reached, and the equilibrium morphologies of double droplets for the seven cases are depicted in the right panel of Fig.~\ref{double_droplet}. It is seen that seven different equilibrium morphologies are exhibited and they can be described as complete engulfing of green fluid by red fluid (i), critical engulfing of green fluid by red fluid (ii), separate dispersion or non-engulfing (iii), kissing (iv), partial engulfing (v), critical engulfing of red fluid by green fluid (vi), and complete engulfing of red fluid by green fluid (vii). These simulation results are consistent with the theoretical predictions by Guzowski et al.\cite{guzowski2012the}.

\subsection{Janus droplet}
\label{sec_Janus}
Among the seven morphologies shown in Fig.~\ref{double_droplet}, the double droplets with partial engulfing morphology are often referred to as the Janus droplet. When the interfacial tension between the constituent fluids is negligibly small, the Janus droplet forms a perfect circle, which is known as perfect Janus droplet (PJD)~\cite{guzowski2012the}. Differentiating from the PJD, the Janus droplet that does not exhibit a perfect circle is called as the general Janus droplet (GJD).

\begin{figure*}
	\centering
	\subfloat[General Janus droplet]{\includegraphics[scale=0.4]{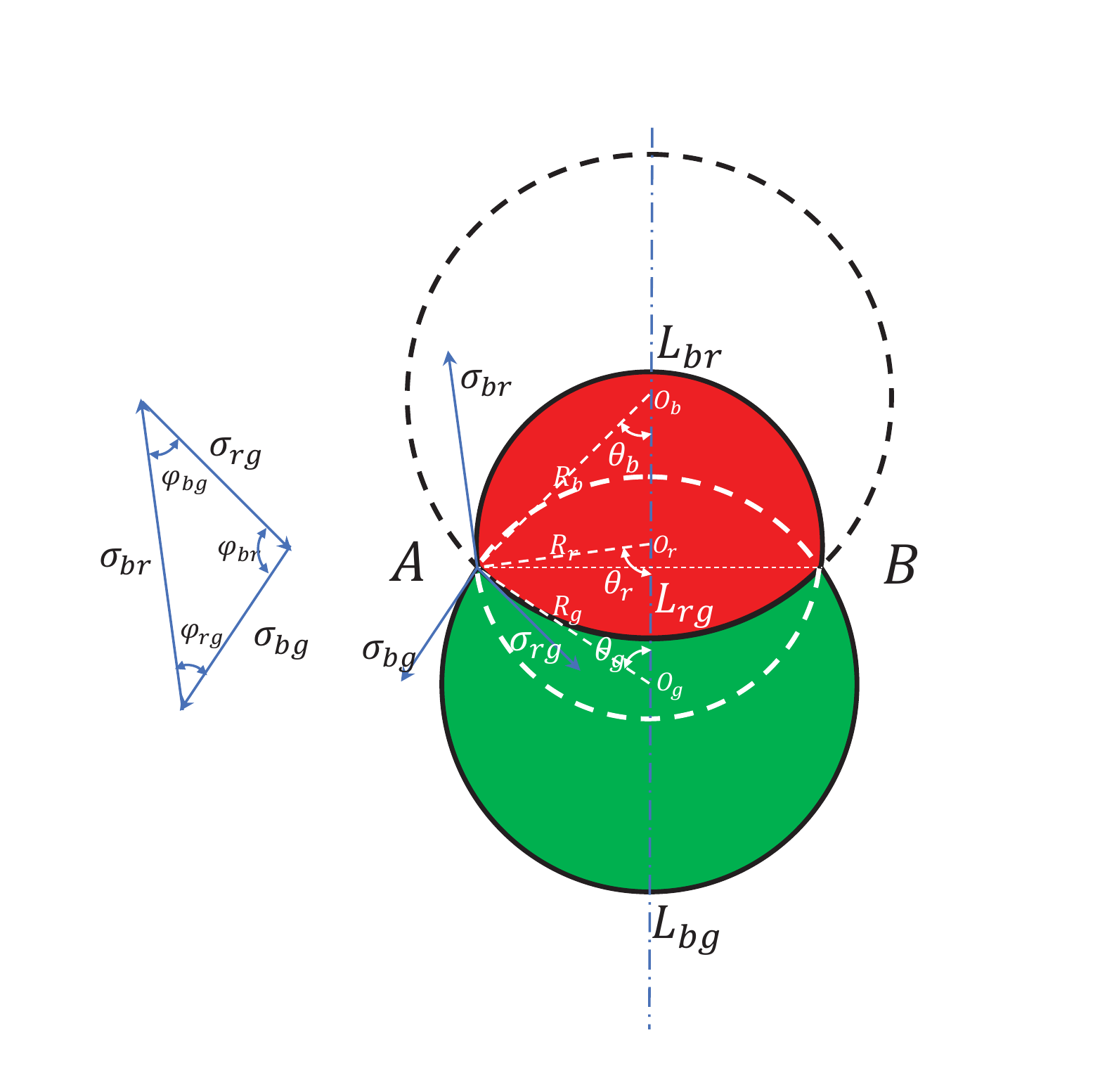}}
	\subfloat[Perfect Janus droplet]{\includegraphics[scale=0.4]{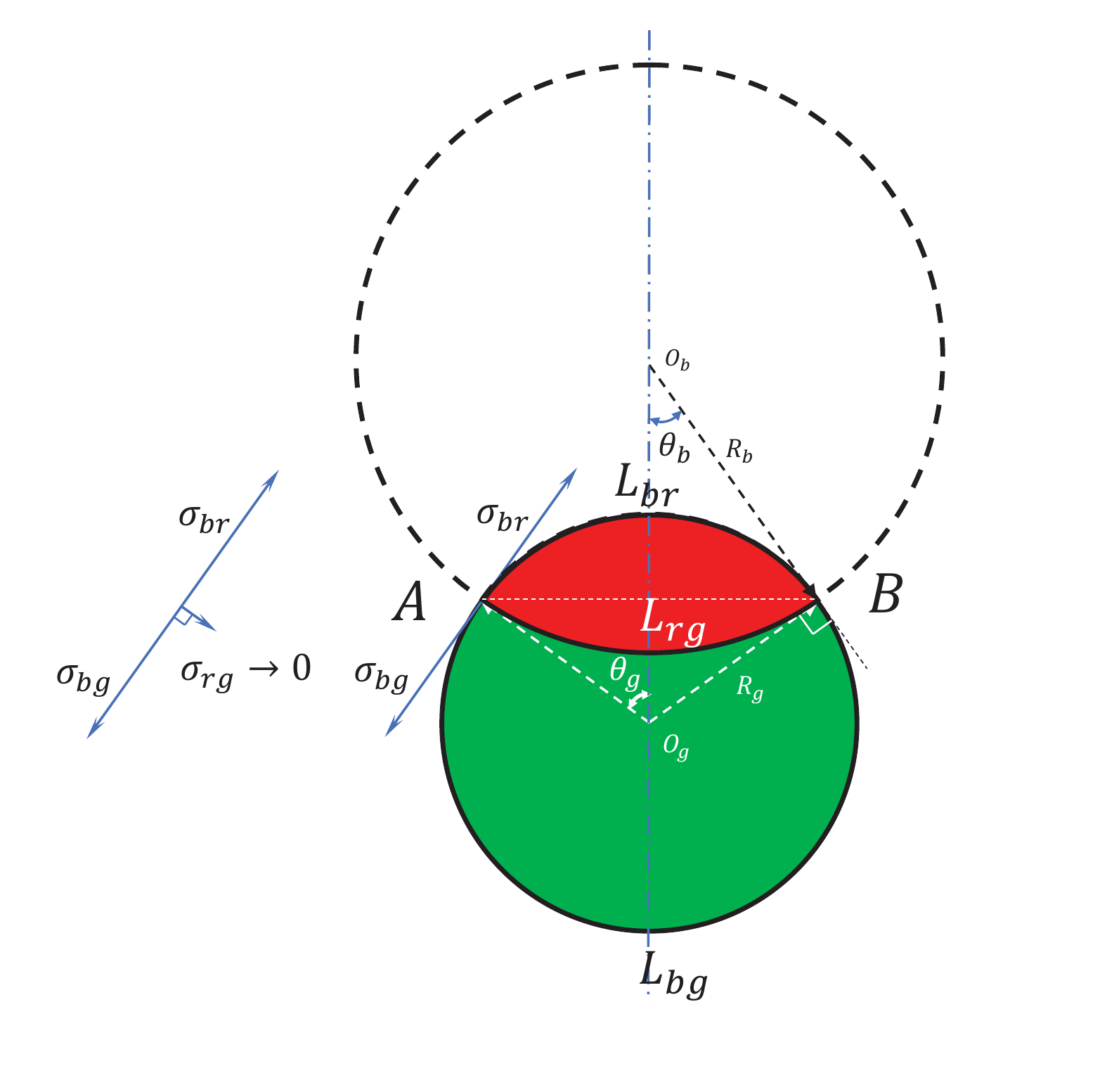}}
	\caption{\ Equilibrium geometry and force balance at a three-phase junction for (a) a general Janus droplet (GJD) and (b) a perfect Janus droplet (PJD).}
	\label{geometry}
\end{figure*}
A Janus droplet, consisting of red and green fluids, is immersed in a static blue fluid. Fig.~\ref{geometry} shows the equilibrium geometries of a GJD and a PJD, as well as the corresponding force balances at three-phase junctions. In this figure, $R_{r}$, $R_{g}$ and $R_{b}$ are the curvature radii of the $rb$, $gb$ and $rg$ interfaces respectively ($kl$ interface refers to the interface between fluid $k$ and fluid $l$); $\theta_{r}$, $\theta_{g}$ and $\theta_{b}$ are the half of the central angles subtended by the chord $AB$; and $d_{rg}$ ($d_{gb}$) is the distance between the centers $O_{g}$ and $O_{r}$ ($O_{b}$). For a GJD, provided that four independent geometric parameters, e.g. $R_{r}$, $R_{g}$, $R_{b}$ and $d_{rg}$, are given, one can analytically obtain all the other geometric parameters, including $d_{gb}$, $\theta_{r}$, $\theta_{g}$ and $\theta_{b}$, and the relative magnitudes of $\sigma_{rb}$, $\sigma_{gb}$ and $\sigma_{rg}$. Specifically, the half of the central angle $\theta_{g}$ can be first calculated by
\begin{eqnarray}\label{eq_gjd_01}
\theta_{g}=\arccos{\frac{R_{g}^{2}+d_{rg}^{2}-R_{r}^{2}}{2R_{g}d_{rg}}},
\end{eqnarray}
which is used to calculate the other two angles, $\theta_{r}$ and $\theta_{b}$, and $R_{b}$ according to
\begin{eqnarray}\label{eq_gjd_02}
R_{g}\sin{\theta_{g}}=R_{b}\sin{\theta_{b}}=R_{r}\sin{\theta_{r}},
\end{eqnarray}
and the distance $d_{gb}$ is then obtained as
\begin{eqnarray}\label{eq_gjd_03}
d_{gb}=R_{g}\cos{\theta_{g}}+R_{b}\cos{\theta_{b}}.
\end{eqnarray}
Next, we determine the angles $\varphi_{rg}$, $\varphi_{rb}$ and $\varphi_{gb}$ through the geometric relationship and the Neumann's triangle shown in Fig.~\ref{geometry}(a). For example, when $R_{g}\cos{\theta_{g}}<{d_{gb}}$ and $R_{g}\cos{\theta_{g}}\geq{d_{rg}}$, these angles can be calculated by
\begin{eqnarray}\label{eq_gjd_04}
\left\{
\begin{aligned}
\varphi_{rb}=\frac{\pi}{2}-\theta_{r}+\theta_{g} \\
\varphi_{gb}=\pi-\theta_{r}-\theta_{b} \\
\varphi_{rg}=\pi-\varphi_{rb}-\varphi_{gb}
\end{aligned}
\right.;
\end{eqnarray}
and on the other hand, when $R_{g}\cos{\theta_{g}}<{d_{gb}}$ and $R_{g}\cos{\theta_{g}}<{d_{rg}}$, we have
\begin{eqnarray}\label{eq_gjd_05}
\left\{
\begin{aligned}
\varphi_{rb}=\theta_{b}+\theta_{g} \\
\varphi_{gb}=\theta_{r}-\theta_{b} \\
\varphi_{rg}=\pi-\varphi_{rb}-\varphi_{gb}
\end{aligned}
\right..
\end{eqnarray}
Finally, one can obtain the relative magnitudes of the interfacial tensions by the law of Sines:
\begin{eqnarray}\label{eq_gjd_06}
\frac{\sigma_{rg}}{\sin{\varphi_{rg}}}=\frac{\sigma_{rb}}{\sin{\varphi_{rb}}}=\frac{\sigma_{gb}}{\sin{\varphi_{gb}}}.
\end{eqnarray}
In other words, all the interfacial tensions can be determined from Eq.(\ref{eq_gjd_06}) if one of them is also given, as we shall do below.

By contrast, the geometry of a PJD is only determined by two areas of the dispersed fluids, i.e. $A_r$ and $A_g$, and its analytical solution is given by
\begin{eqnarray}\label{eq_pjd}
\left\{
\begin{aligned}
R_{g}=R_{r}=\sqrt{\frac{A_{g}+A_{r}}{\pi}} \\
R_{b}=R_{g}\tan{\theta_{g}} \\
\frac{A_{r}}{A_{r}+A_{g}}=\frac{\theta_{g}-\sin{\theta_{g}}\cos{\theta_{g}}+\tan^{2}{\theta_{g}}\left(\frac{\pi}{2}-\theta_{g}-\sin{\theta_{g}}\cos{\theta_{g}}\right)}{\pi} \\
d_{gb}=R_{g}\cos{\theta_{g}}+R_{b}\cos{\theta_{b}}
\end{aligned}
\right..
\end{eqnarray}
The above equation suggests that one can obtain all the other geometric parameters, such as $R_{b}$, $\theta_{g}$ and $d_{gb}$, if the area ratio $\frac{A_{r}}{A_{r}+A_{g}}$ and $R_{g}$ are given.

To test the accuracy of the present model for Janus droplets, we conduct two groups of simulations with one for GJD and the other for PJD. The size of the computational domain is set as $[1,300]\times[1,300]$, and the periodic boundary conditions are used at all the boundaries. The kinematic viscosities for all the fluids are fixed at $\nu_{k}=0.1$. In the GJD simulations, we select $\sigma_{rg}=0.01$, $R_{r}=60$, $R_{g}=80$ and $R_{b}=160$, and vary the distance $d_{rg}$ from 40 to 120 with an increment of 20. Using these parameters, we can compute the geometric parameters $d_{rg}$ and $d_{gb}$ as well as the interfacial tensions $\sigma_{gb}$ and $\sigma_{rg}$ through Eqs.~(\ref{eq_gjd_01}) to (\ref{eq_gjd_06}), which are presented in Table~\ref{tab_gjd}. We initialize the fluid distribution such that it follows the given and analytically computed geometric parameters described above, and assume that the circles for $gb$, $rb$, and $rg$ interfaces are initially centered at $(150.5,R_{g}+10)$, $(150.5,R_{g}+10+d_{rg})$ and $(150.5,R_{g}+10+d_{gb})$, respectively. 
\begin{table*}[!htb]
	\small
	\centering
	\caption{\ The interfacial tensions $\sigma_{gb}$ and $\sigma_{rb}$ and the distance $d_{gb}$ calculated from Eqs.~(\ref{eq_gjd_01})-(\ref{eq_gjd_06}) for GJDs with $\sigma_{rg}=0.01$, $R_{r}=60$, $R_{g}=80$ and $R_{b}=160$ at different values of $d_{rg}$.}
	\label{tab_gjd}
	\begin{tabular*}{\textwidth}{@{\extracolsep{\fill}}llll}
		\hline
		$d_{rg}$&$\sigma_{gb}$&$\sigma_{rb}$&$d_{gb}$\\
		\hline
		20 & 0.01 & 0.02 & -- \\
		40 & 0.01699 & 0.01497 & 204.08 \\
		60 & 0.01182 & 0.01261 & 201.81 \\
		80 & 0.00797 & 0.00973 & 207.52 \\
		100 & 0.00583 & 0.00812 & 216.63 \\
		120 & 0.00449 & 0.00712 & 227.67 \\
		140 & 0.00357 & 0.00643 & 240.00 \\
		\hline
	\end{tabular*}
\end{table*}
In the PJD simulations, we select $\sigma_{rb}=\sigma_{gb}=0.01$, $\sigma_{rg}=1\times{10^{-8}}$ and $R_{r}=R_{g}=80$, and vary the area fraction $\frac{A_{r}}{A_{r}+A_{g}}$ from 0.1 to 0.5 with an increment of 0.1. These parameters allow us to analytically compute all the other geometric parameters of a PJD, e.g. $R_b$ and $d_{gb}$, which are listed in Table \ref{tab_pjd}. We follow the analytical geometric parameters to initialize the fluid distribution, and assume that the circles for $gb$, $rb$ and $rg$ interfaces are initially located at $(150.5,150.5)$, $(150.5,150.5)$ and $(150.5,150.5+d_{gb})$, respectively. In particular, we note that $\frac{A_{r}}{A_{r}+A_{g}}=0.5$ leads to $R_{b}\rightarrow\infty$ and $d_{gb}\rightarrow\infty$, suggesting that the interface $rg$ is theoretically a straight line located at $y=150.5$.
\begin{table*}[!htb]
	\small
	\centering
	\caption{\ The geometric parameters $R_b$ and $d_{gb}$ calculated from Eq.(\ref{eq_pjd}) for PJDs with $R_{r}=R_{g}=80$ at different area fractions.}
	\label{tab_pjd}
	\begin{tabular*}{\textwidth}{@{\extracolsep{\fill}}llllll}
		\hline
		$\frac{A_{r}}{A_{r}+A_{g}}$ &0.1 &0.2 &0.3 &0.4 &0.5\\
		\hline
		$R_{b}$ & 48.32 & 88.39 & 154.12 & 331.92 & $\infty$\\
		\hline
		$d_{gb}$ & 93.46 & 119.22 & 173.64 & 341.43 & $\infty$\\
		\hline
	\end{tabular*}
\end{table*}

\begin{figure*}[!htb]
	\centering
	\subfloat[$d_{rg}=40$]{\includegraphics[scale=0.22]{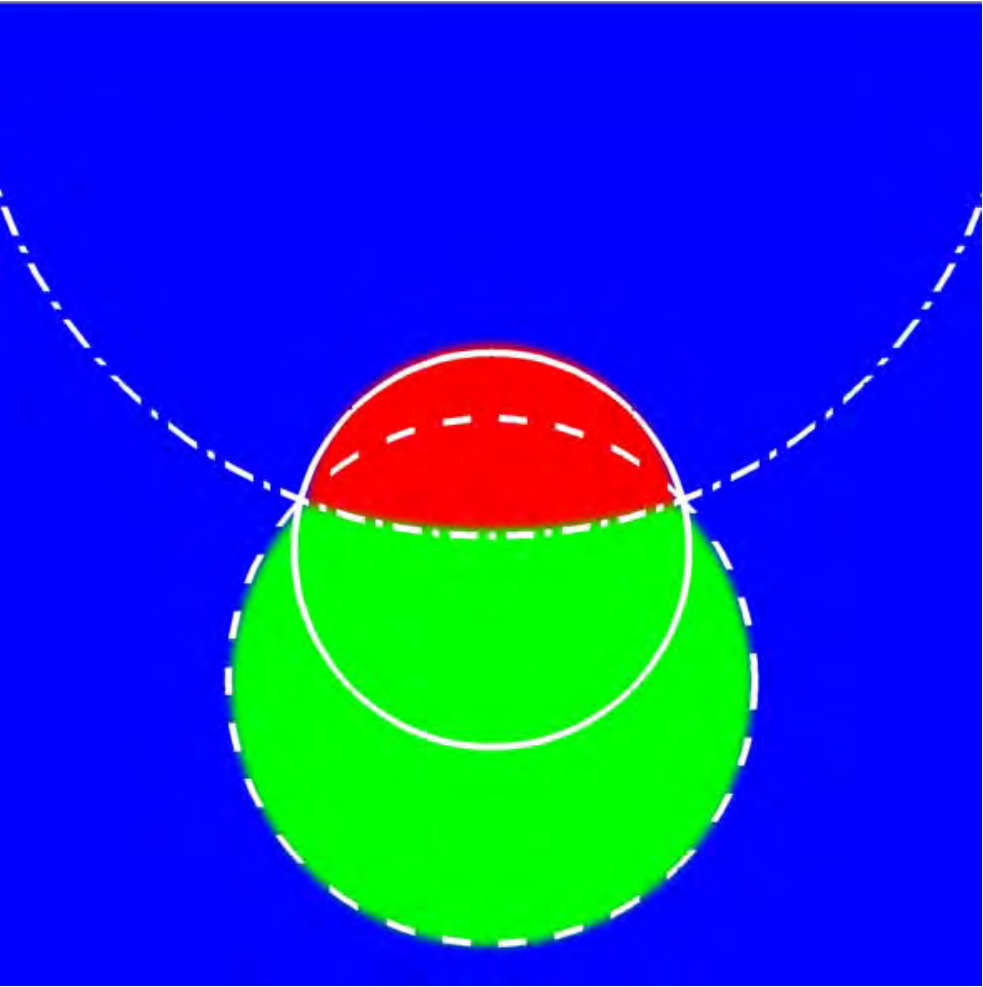}}
	\subfloat[$d_{rg}=60$]{\includegraphics[scale=0.22]{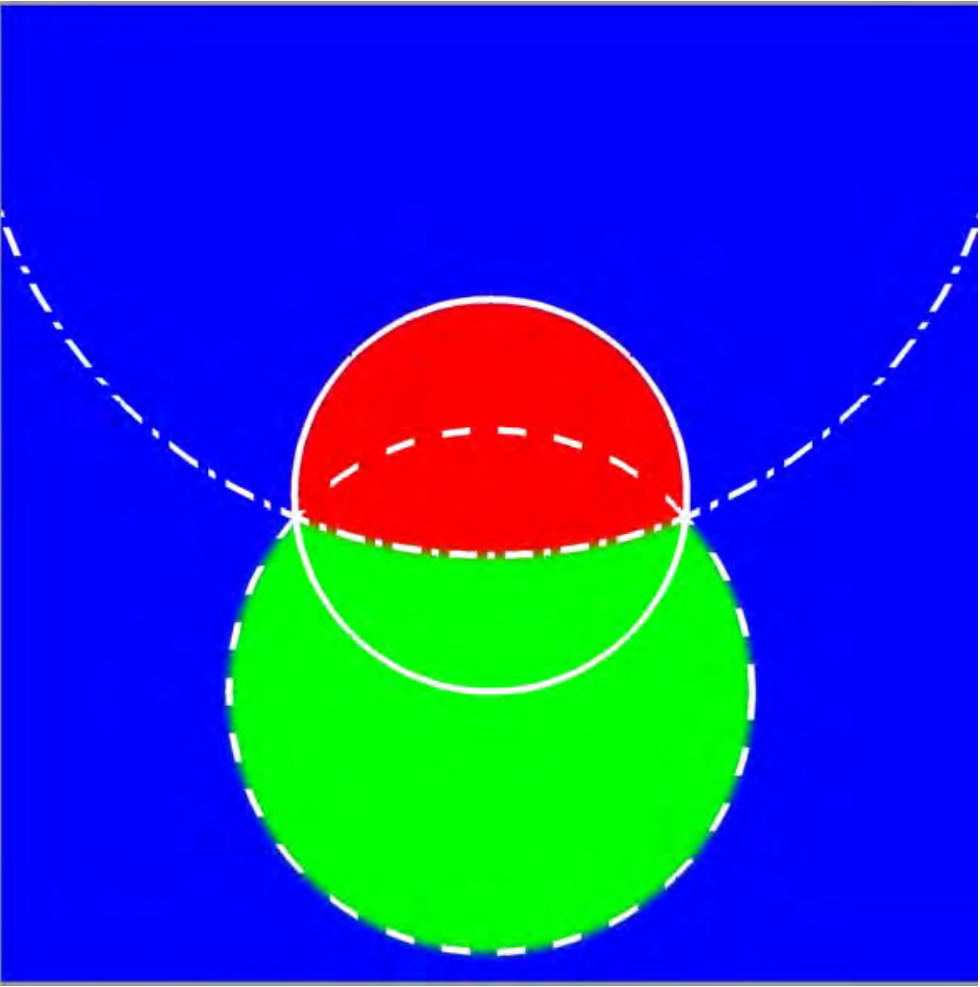}}
	\subfloat[$d_{rg}=80$]{\includegraphics[scale=0.22]{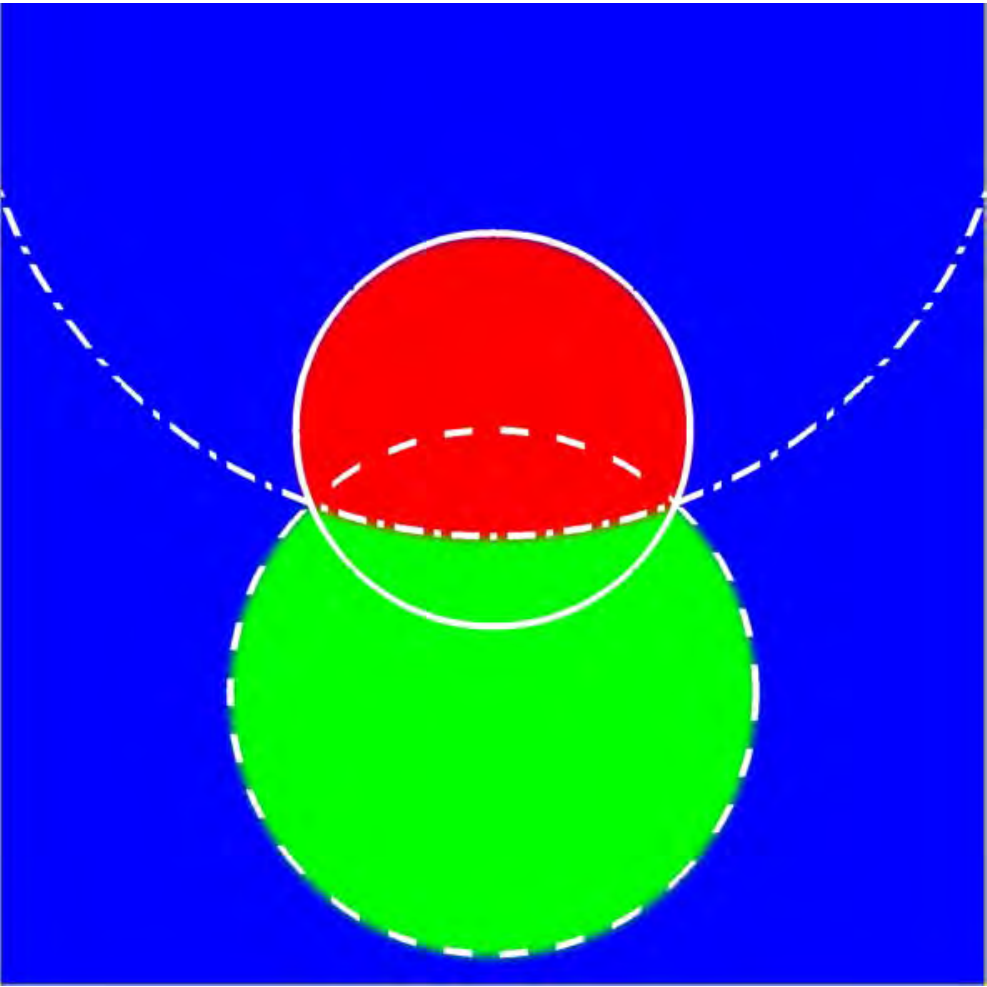}}
	\subfloat[$d_{rg}=100$]{\includegraphics[scale=0.22]{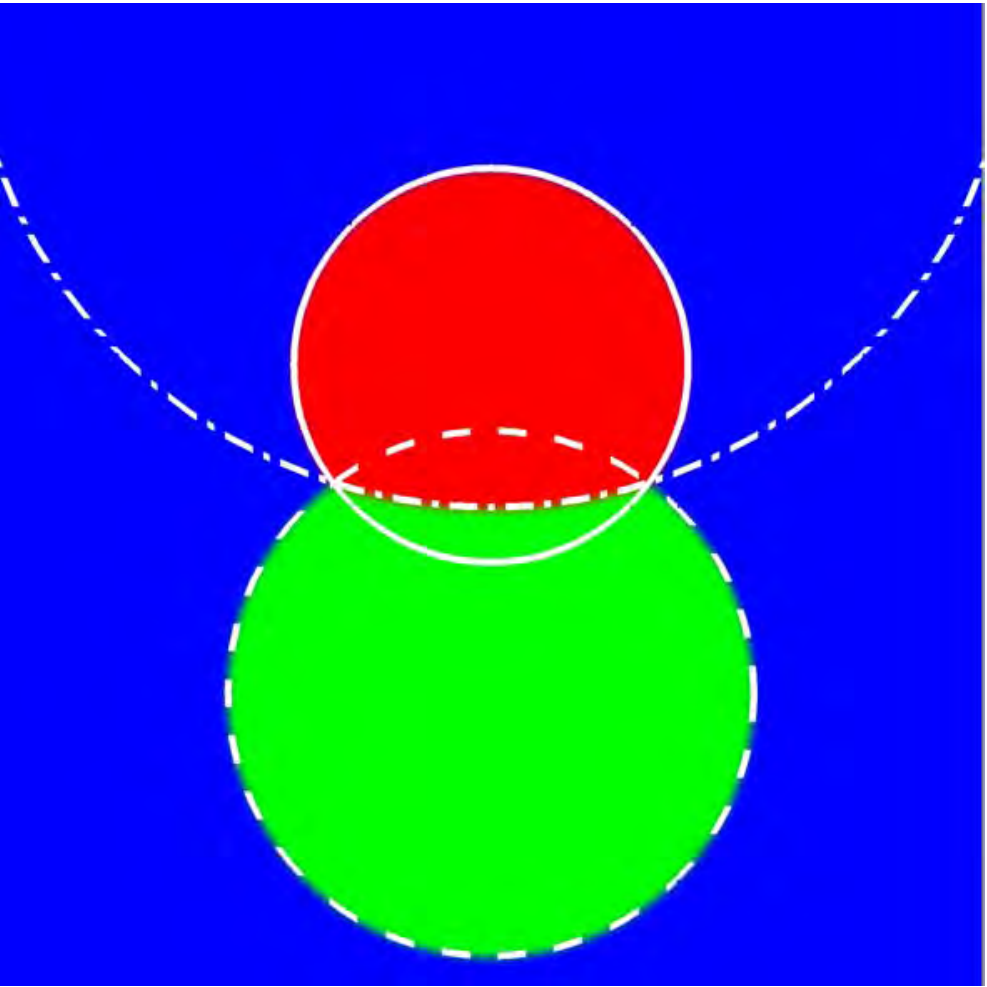}}
	\subfloat[$d_{rg}=120$]{\includegraphics[scale=0.22]{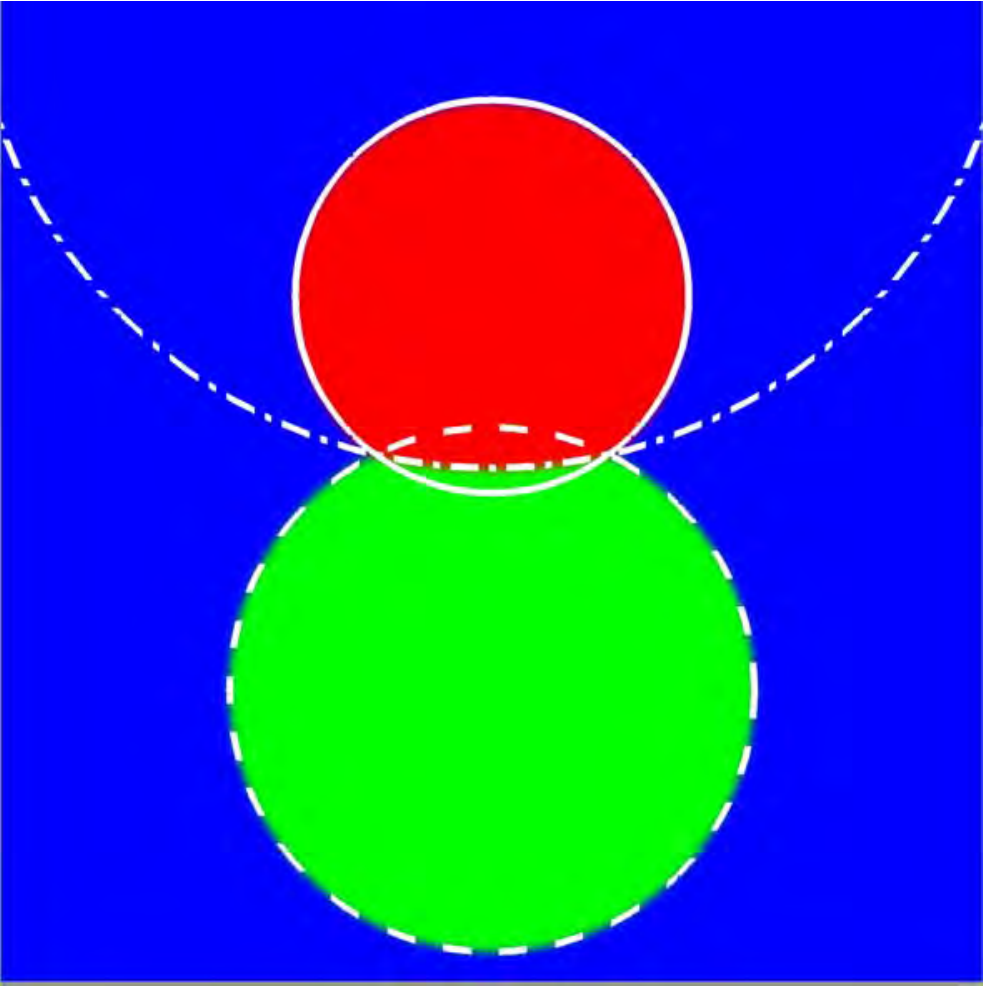}}
	\caption{\ Comparison between the analytical and simulated results for GJDs with $\sigma_{rg}=0.01$, $R_{r}=60$, $R_{g}=80$ and $R_{b}=160$ at different values of $d_{rg}$. The analytical interface profiles are represented by the white lines of different patterns, while the simulated red, green and blue fluids are indicated in red, green and blue, respectively.}
	\label{general_ball}
\end{figure*}

\begin{figure*}[!htb]
	\centering
	\subfloat[$\frac{A_{r}}{A_{r}+A_{g}}=0.1$]{\includegraphics[scale=0.22]{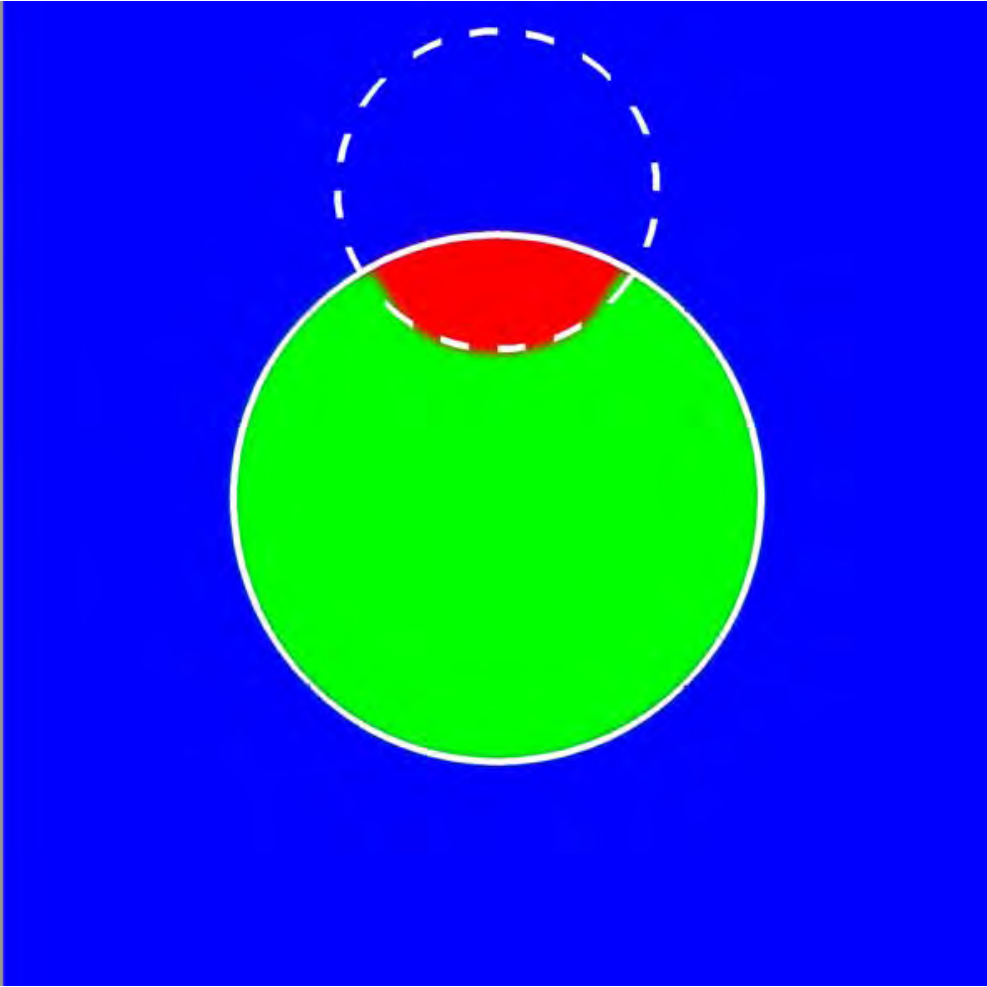}}
	\subfloat[$\frac{A_{r}}{A_{r}+A_{g}}=0.2$]{\includegraphics[scale=0.22]{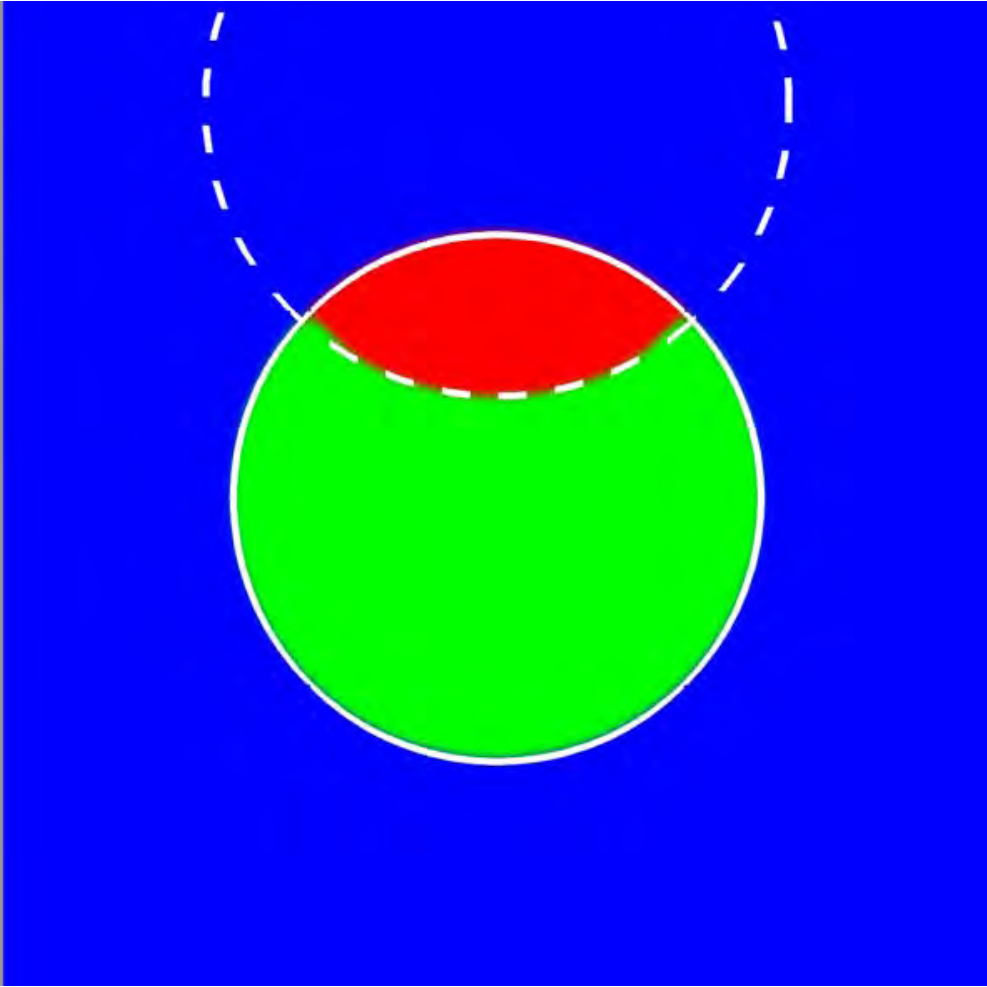}}
	\subfloat[$\frac{A_{r}}{A_{r}+A_{g}}=0.3$]{\includegraphics[scale=0.22]{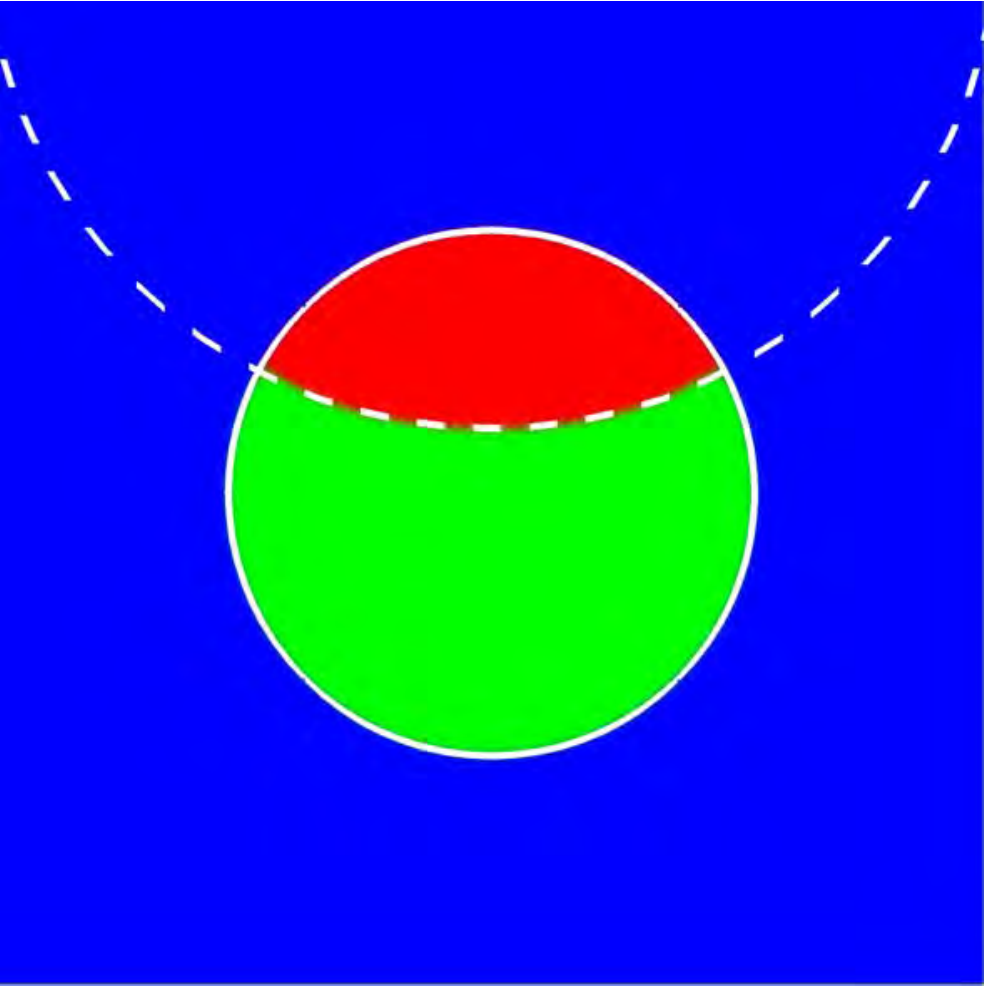}}
	\subfloat[$\frac{A_{r}}{A_{r}+A_{g}}=0.4$]{\includegraphics[scale=0.22]{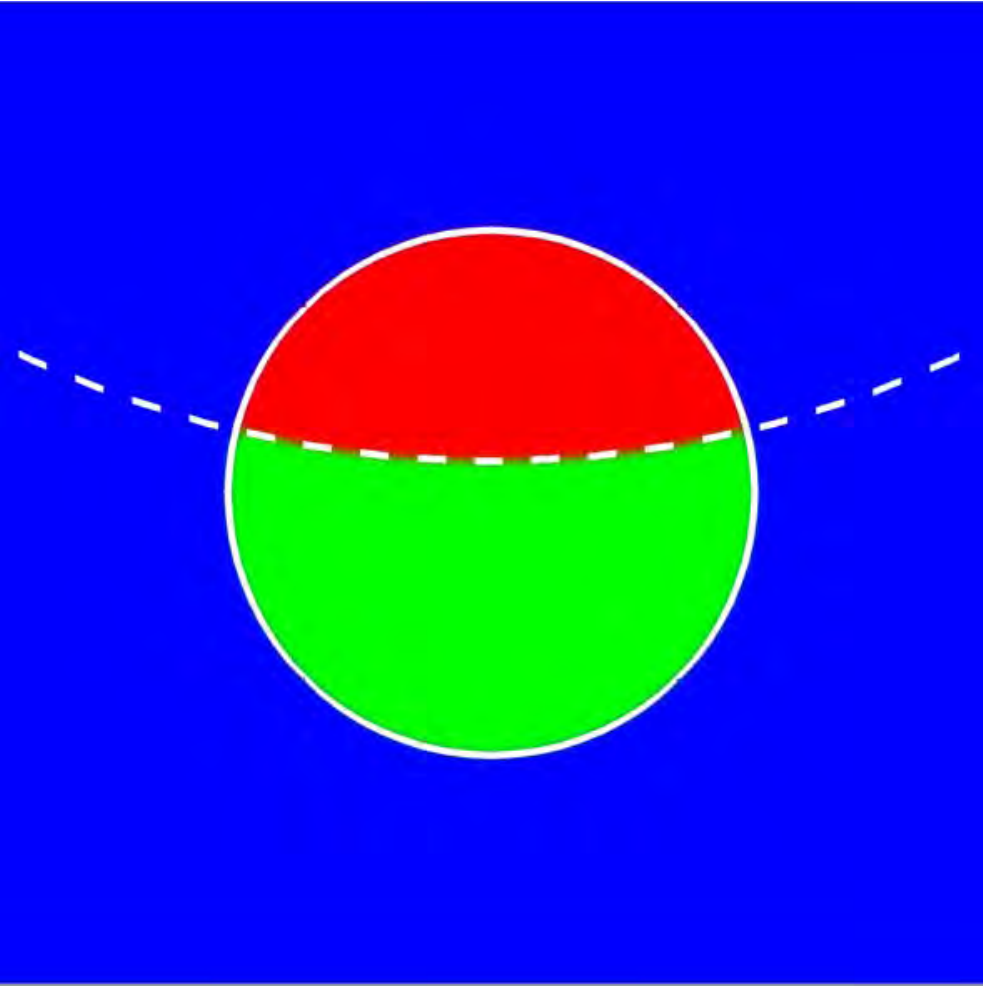}}
	\subfloat[$\frac{A_{r}}{A_{r}+A_{g}}=0.5$]{\includegraphics[scale=0.22]{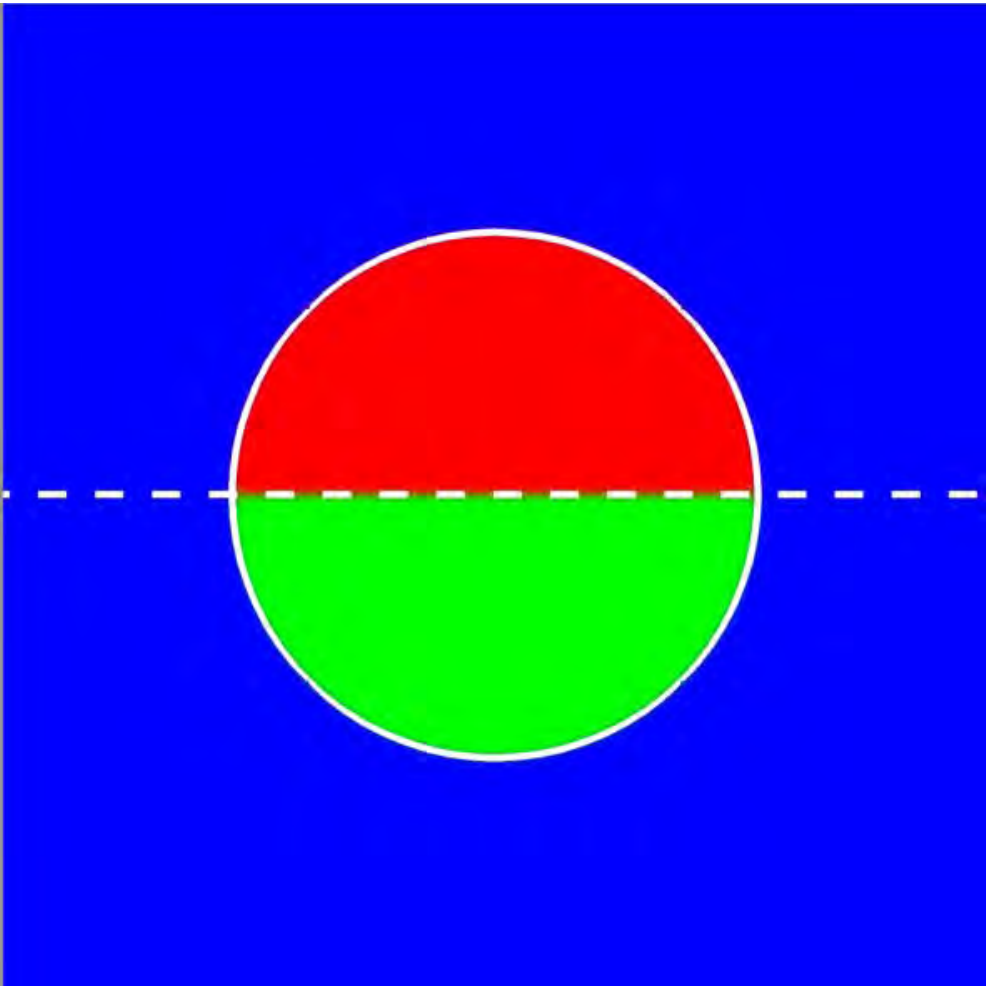}}
	\caption{\ Comparison between the analytical and simulated results for PJDs with $R_{r}=R_{g}=80$ at different area fractions. The analytical interface profiles are represented by the white lines of different patterns, while the simulated red, green and blue fluids are indicated in red, green and blue, respectively.}
	\label{perfect_ball}
\end{figure*}

All of the simulations are run until a steady state is reached. Figs.~\ref{general_ball} and \ref{perfect_ball} show the comparison between the analytical and simulated results for the GJDs and PJDs. In each of the figures, the analytical interface profiles are represented by the white lines of different patterns, while the red, green and blue fluids are indicated in red, green and blue, respectively. It is seen that our simulation results agree well with the analytical ones for various geometry configurations of GJD and PJD. 

\subsection{Near-critical and critical states}
For double droplets immersed in a static matrix, the critical state occurs when the largest interfacial tension equals the sum of the other two. As previously shown in Fig.~\ref{double_droplet}, the critical state of double droplets can be subdivided into the kissing state ((iv) in Fig.~\ref{double_droplet}) and the critical engulfing state ((ii) or (vi) in Fig.~\ref{double_droplet}). For the convenience of description, we define the near-critical state as the state where the largest interfacial tension is close to the sum of the other two. It is challenging to accurately simulate the critical and near-critical states, where a slight inaccuracy in modeling could lead to significant simulation errors.

To highlight the strength of the present model for critical scenarios, we consider the kissing/near-kissing states and the critical/near-critical engulfing states in a rectangular domain of $[1,300]\times[1,300]$. The boundary conditions and fluid viscosities are set the same as those in Section~\ref{sec_Janus}. In the kissing/near-kissing states, a pair of equal-sized droplets with the radii of $R_{r}=R_{g}=60$ are initially placed with a distance of $d_{rg}$, and they are symmetric with respect to the centerline $y=150.5$. The simulations are performed for a constant $\sigma_{rg}$ of $0.01$ but varying $\sigma_{gb}$ ($=\sigma_{rb}$), which is varied around the critical value of $0.005$ with an increment of $2\times 10^{-4}$. Note that the initial distance $d_{rg}$ depends on the value of $\sigma_{gb}$, and is given by its analytical value in equilibrium as
\begin{eqnarray}\label{eq_drg}
d_{rg}=\left\{
\begin{aligned}
&R_{r}\sigma_{rg}/\sigma_{gb} & \mathrm{if}~ \sigma_{gb}>0.005;\\
&R_{r}+R_{g}=120 & \mathrm{otherwise}.
\end{aligned}
\right.
\end{eqnarray}

In the critical/near-critical engulfing states, we consider a green droplet with $R_{g}=80$ entirely or partially engulfing a red droplet with $R_{r}=60$ for $\sigma_{gb}=\sigma_{rg}=0.01$. $\sigma_{rb}$ is varied around the critical value of 0.02 with an increment of $2\times 10^{-4}$. With these parameters, we are able to analytically compute other geometric parameters, which are given by  $d_{rg}=\sqrt{R_{r}^{2}+R_{g}^{2}-2R_{r}R_{g}\cos{\alpha}}$, $R_{b}=\frac{R_{g}\sin{\theta_{g}}}{\sin{\theta_{b}}}$ and $d_{gb}=R_{g}\cos{\theta_{g}}-R_{b}\cos{\theta_{b}}$ for $\sigma_{rb}<0.02$, and by $d_{rg}=R_{g}-R_{r}=20$ for $\sigma_{rb}\ge 0.02$.
Herein, $\cos{\alpha}=\frac{\sigma_{rb}}{2\sigma_{gb}}$, $\theta_{g}=\arccos{\frac{R_{g}^{2}+d_{gr}^{2}-R_{r}^{2}}{2R_{g}d_{gr}}}$ and $\theta_{b}=\theta_{g}+2\alpha$. Again, we initialize the fluid distribution such that it follows the analytical geometric parameters.

\begin{figure*}[!htb]
	\centering
	\subfloat[]{\includegraphics[scale=0.3]{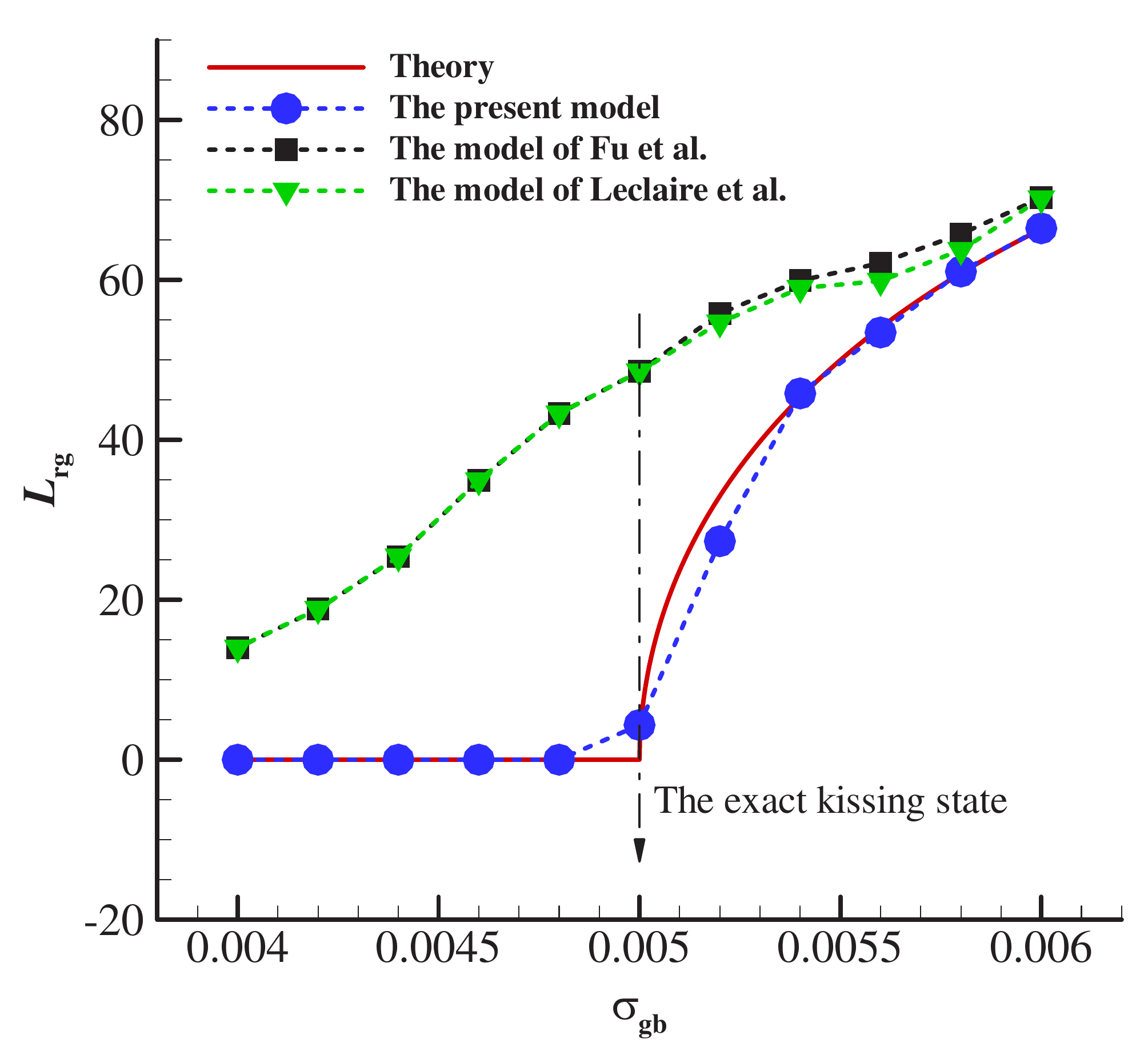}} 
	\subfloat[]{\includegraphics[scale=0.3]{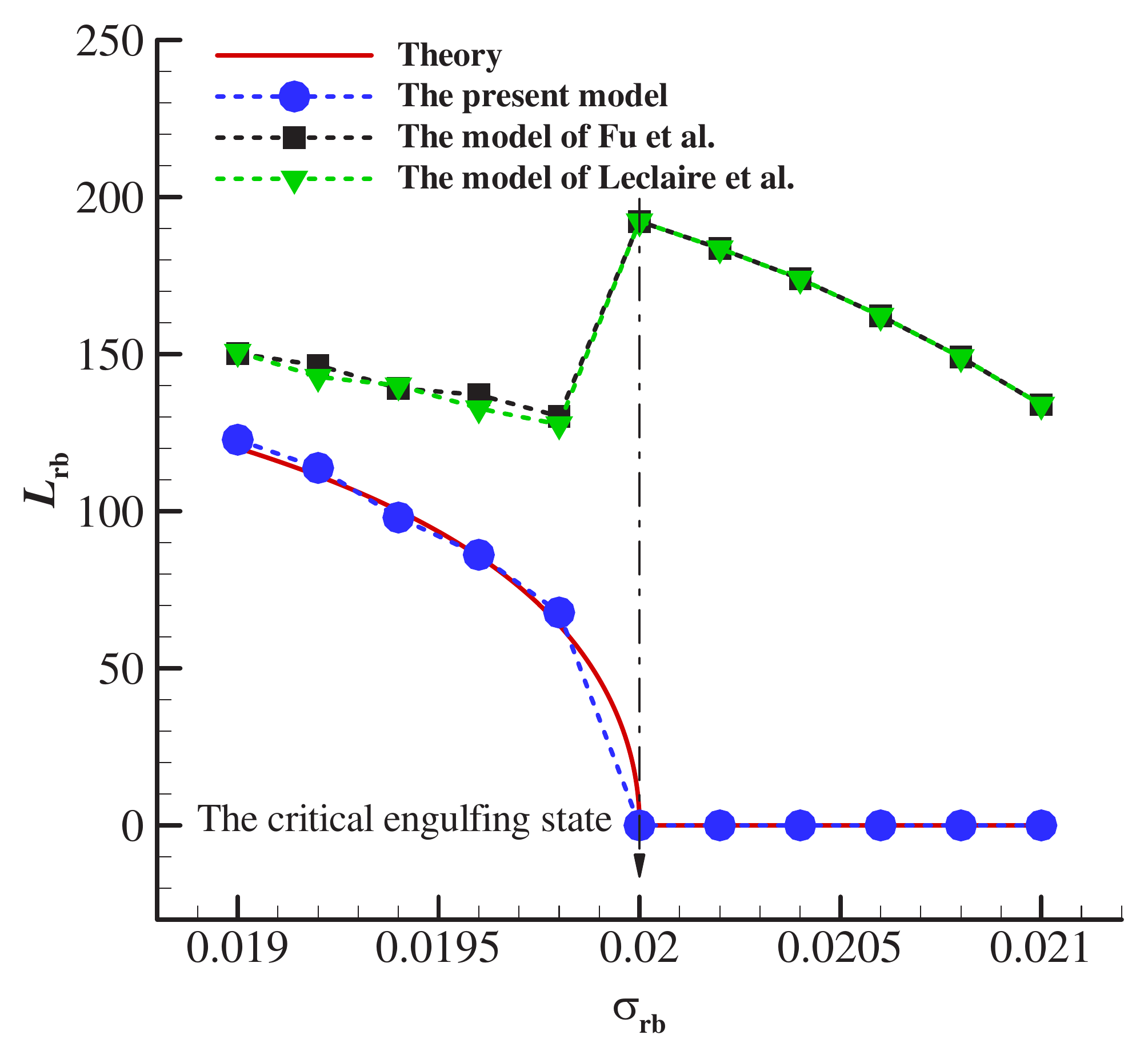}}
	\caption{\ (a) The interface length $L_{rg}$ as a function of $\sigma_{gb}$ in the kissing/near-kissing states; (b) the interface length $L_{rb}$ as a function of $\sigma_{rb}$ in the critical/near-critical engulfing states.}
	\label{fig_critical_state}
\end{figure*}

\begin{figure*}[!thb]
	\centering
	\subfloat[]{\includegraphics[scale=0.3]{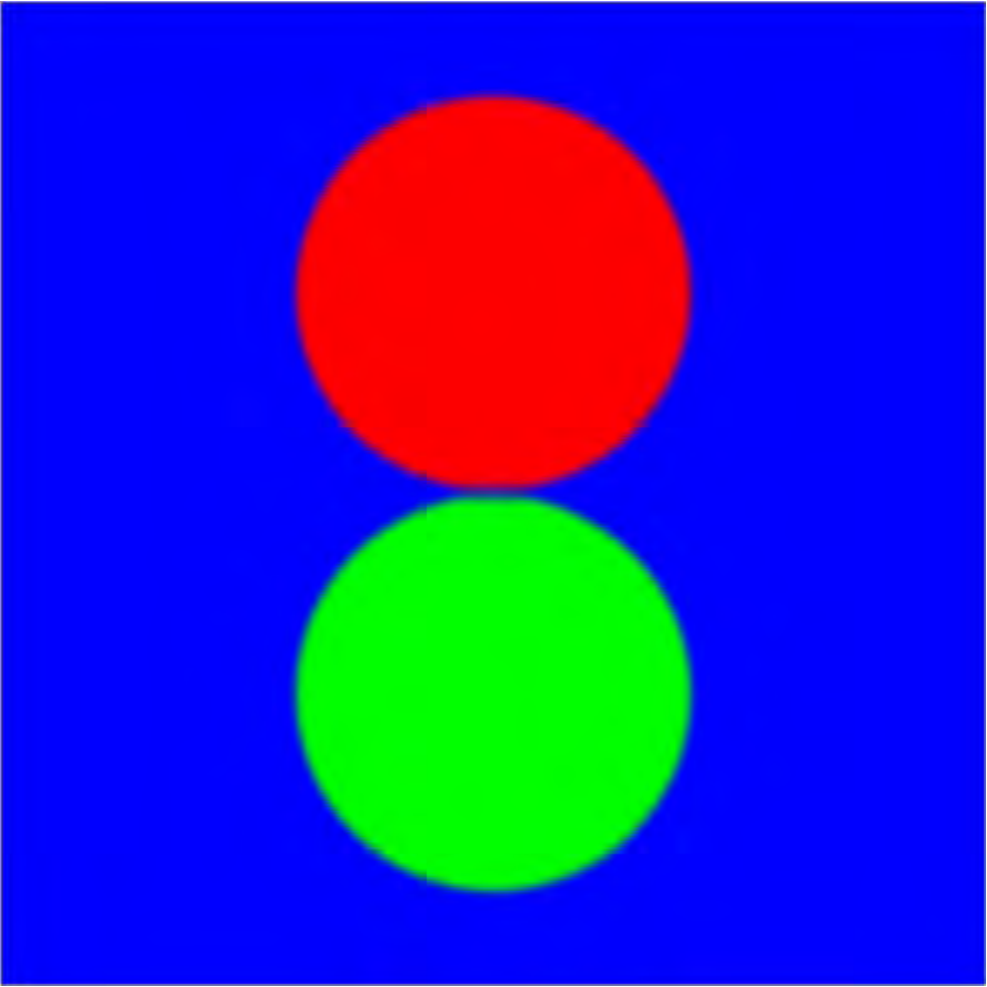}}
	\subfloat[]{\includegraphics[scale=0.3]{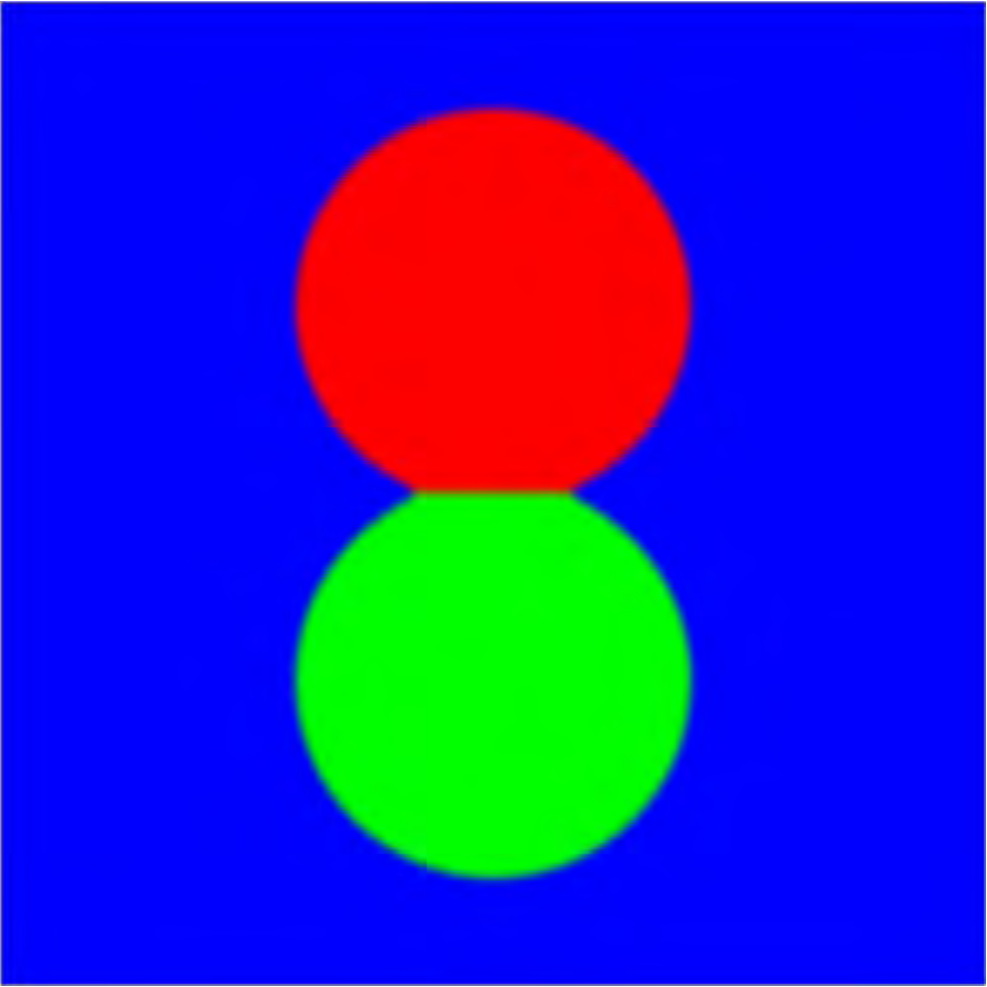}} 
	\subfloat[]{\includegraphics[scale=0.3]{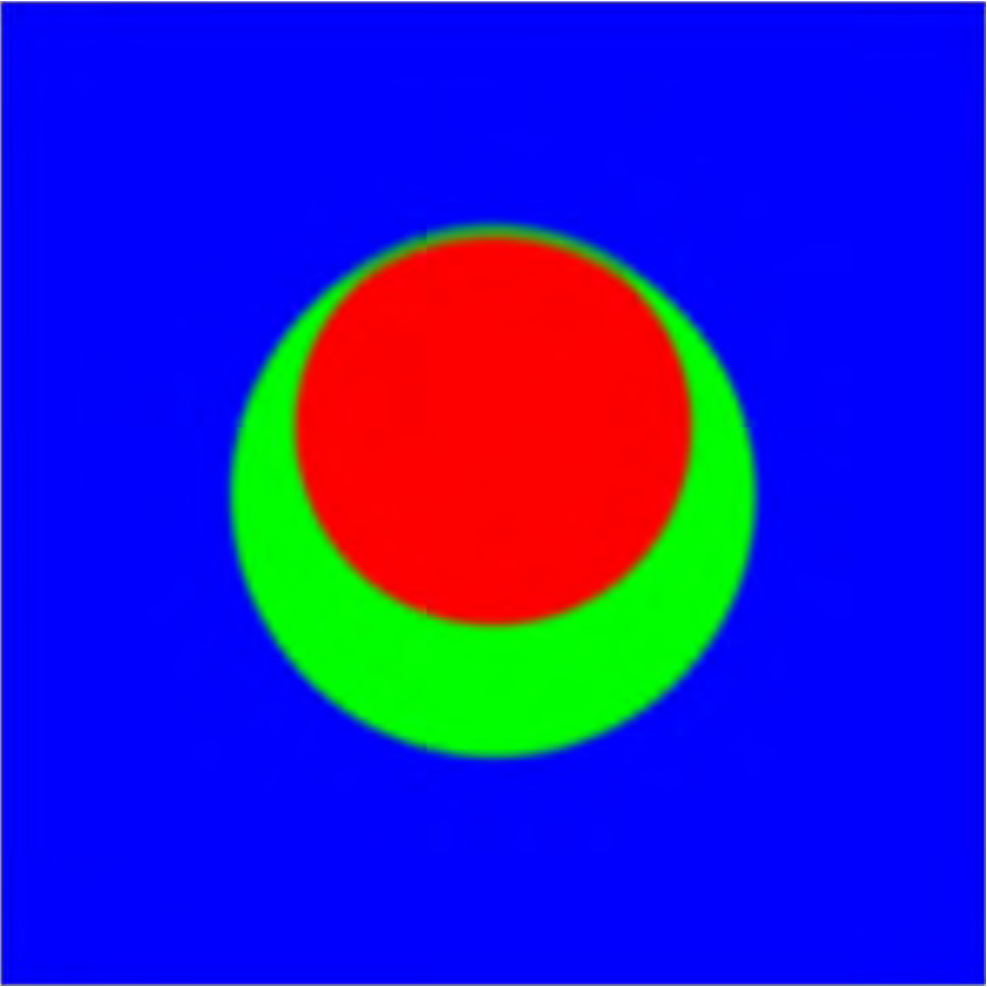}}
	\subfloat[]{\includegraphics[scale=0.3]{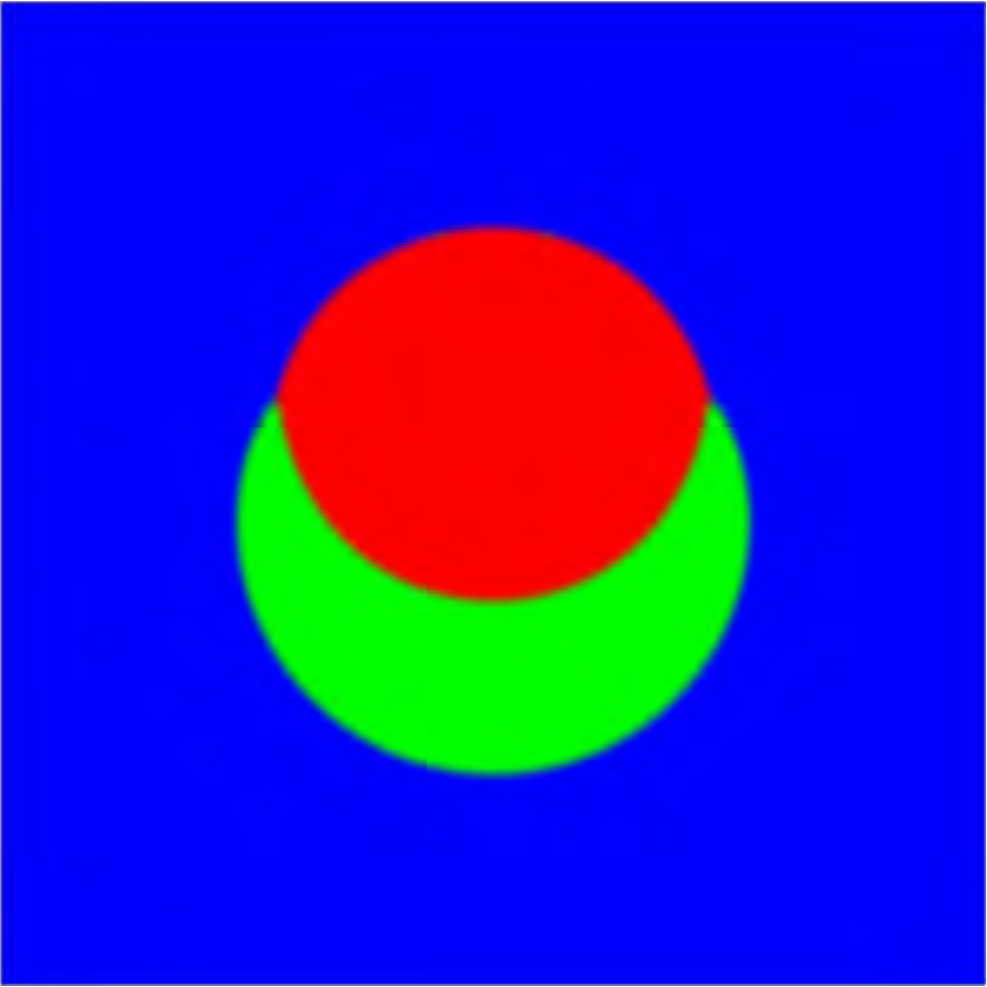}}
	\caption{\ The final fluid distributions obtained by (a) the present model and (b) the model of Fu et al.~\cite{fu2016numerical} for the kissing state, and by (c) the present model and (d) the model of Fu et al.~\cite{fu2016numerical} for the critical engulfing state.}
	\label{critical_compare}
\end{figure*}

In addition to the present model, we also use the model of Fu et al.\cite{fu2016numerical} and the model of Leclaire et al.\cite{leclaire2013progress} for the simulations. When the simulations reach the steady state, we quantify the interface lengths $L_{rg}$ in the kissing/near-kissing states and $L_{rb}$ in the critical/near-critical engulfing states. Fig.~\ref{fig_critical_state} compares the simulated results from the present model with those from the model of Fu et al.\cite{fu2016numerical} and the model of Leclaire et al.\cite{leclaire2013progress}, and the analytical solutions. It is seen that for either kissing/near-kissing states or critical/near-critical engulfing states, the simulated results from the present model are in good agreement with the analytical solutions, while the simulated results from the other two models significantly deviate from the analytical solutions. Fig.~\ref{critical_compare} shows the final fluid distributions obtained by the present model and the model of Fu et al.\cite{fu2016numerical}, in both kissing and critical engulfing states. Note that the fluid distribution from the model of Leclaire et al.\cite{leclaire2013progress} is not shown in the figure, since it produces almost the same results as the model of Fu et al.\cite{fu2016numerical}. Clearly, both critical states are correctly reproduced by the present model but not by the model of Fu et al.\cite{fu2016numerical}. These results indicate that the present model is advantageous to simulate critical state in ternary fluids.

\section{conclusions}
A LB color-gradient model is proposed to simulate immiscible ternary fluids with a full range of interfacial tensions. An interfacial force formulation for $N$-phase ($N\ge 3$) systems is derived and then introduced into the model using a body force scheme, which is found to effectively reduce spurious velocities. A recoloring algorithm proposed by Spencer et al.\cite{spencer2010lattice} is applied to produce the phase segregation and ensure the immiscibility of three different fluids, where a novel form of segregation parameters is proposed by considering the existence of Neumann's triangle and the effect of equilibrium contact angle in three-phase junction. The model's capability in capturing interfaces and modeling interfacial tensions is first validated by the simulation of the two separate static droplets and the Young-Laplace test for a compound droplet. The overall performance of the model is then assessed by simulating the spreading of a droplet between two stratified fluids, and both the partial and complete spreadings are predicted with satisfactory accuracy.

Finally, the present model is used to study the stability and structure of double droplets in a static matrix over a wide range of interfacial tensions. By changing two ratios of the interfacial tensions, seven possible equilibrium morphologies are successfully reproduced, which are consistent with the theoretical stability diagram by Guzowski et al.\cite{guzowski2012the}. For various geometry configurations of general and perfect Janus droplets, good agreemento between simulated results and analytical solutions shows the present model is accurate when three interfacial tensions yield a Neumann's triangle. In addition, we also simulate the near-critical and critical states of double droplets, which is challenging since the outcomes are very sensitive to the model accuracy. It is found that the simulated results from the present model agree well with the analytical solutions, while the simulated results from the existing color-gradient models significantly deviate from the analytical solutions, especially in critical states. In summary, the present work provides the first LB multiphase model that allows for accurate simulation of ternary fluid flows with a full range of interfacial tensions.

\section*{Acknowledgements}
This work was supported by the National Natural Science Foundation of China (Nos. 51506168, 51711530130), the National Key Research and Development Project of China (No. 2016YFB0200902), the China Postdoctoral Science Foundation (No. 2016M590943), Guangdong Provincial Key Laboratory of Fire Science and Technology (No. 2010A060801010) and Guangdong Provincial Scientific and Technological Project (No. 2011B090400518). Y. Yu was supported by the China Scholarship Council for one year study at the University of Strathclyde, UK. H. Liu gratefully acknowledges the financial supports from Thousand Youth Talents Program for Distinguished Young Scholars, the Young Talent Support Plan of Xi'an Jiaotong University. 

\bibliographystyle{model1-num-names}
\bibliography{refs}

\end{document}